\documentclass[12pt,useAMS]{ociamthesis}  % default square logo 
\usepackage{amssymb}
\usepackage{footnote, enumerate, url, amsmath, color}
\usepackage{multirow}
\usepackage{tabularx}
\usepackage{amssymb}
\usepackage{graphicx}
\usepackage{epsfig}
\usepackage{psfrag}
\usepackage{dcolumn}
\usepackage{bm}
\usepackage{amsmath,amssymb,stmaryrd} 
\usepackage{bbm}
\usepackage{physics}
\usepackage{pdfpages}

\definecolor{titlecol}{rgb}{0,0,0}
\definecolor{titlecol2}{rgb}{0,0.65,0}
\definecolor{titlecol3}{rgb}{0.99,0.4,0.}
\definecolor{minorcol}{rgb}{0,0,0}
\def\lesssim{\mathrel{\hbox{\rlap{\hbox{\lower3pt\hbox{$\sim$}}}\hbox{\raise2pt\hbox{$<$}}}}}
\def\gtrsim{\mathrel{\hbox{\rlap{\hbox{\lower3pt\hbox{$\sim$}}}\hbox{\raise2pt\hbox{$>$}}}}}

\newcommand\id{\ensuremath{\mathbbm{1}}}

\title{Light-matter interaction in 2D materials in weak and strong-coupling regimes}   %note \\[1ex] is a line break in the title

\author{Dogyun Ko}             %your name
\college{Center for Theoretical Physics of Complex Systems, IBS School University of Science and Technology}  %your college

\degree{Doctor of Philosophy}     %the degree
\degreedate{August 2023}         %the degree date

\begin{document}
\setcounter{secnumdepth}{3}
\setcounter{tocdepth}{3}

\maketitle

\vspace*{3cm}

% \includepdf[pages={1}]{add_pages/3.pdf}
% \includepdf[pages={2}]{add_pages/4.pdf}
% \includepdf[pages={1}]{add_pages/cover.pdf}
% \includepdf[pages={1}]{super.pdf}
% \includepdf[pages={1}]{sig_sig.pdf}
\begin{acknowledgements}

I'm deeply grateful to my supervisor Prof. Ivan Savenko for giving me the excellent opportunity to perform doctoral research in his group and for his support and expert advice. I would also like to thank my colleague Dr. Meng Sun for the helpful discussions for the entire projects we worked on together. I also thank Prof. Vadim Kovalev, Prof. Yuri Rubo, Prof. Alexei Andreanov, Dr. Anton Parafilo, and Dr. Ihor Vakulchyk for valuable discussions. I am also grateful to my collaborators at KAIST, in particular, Dr. Daegwang Choi, Prof. Hyoungsoon Choi, and Prof. Yong-Hoon Cho for their fruitful cooperation with our team.

\end{acknowledgements}

\begin{abstract}
This thesis studies light-matter interactions in strong and weak coupling regimes. 
In the first part, 
we study the formation and propagation of exciton-polariton condensates in different microcavities in the strong coupling regime. 
Exciton-polaritons are composite quasiparticles created as a result of the strong coupling between microcavity photons and quantum well excitons. 

In the first part of the thesis, we take a system of exciton-polaritons in a Kagome lattice and show that an initially localized condensate propagates in a specific direction in space in the presence of anisotropy in the lattice, and the initially localized condensate experiences revivals. We also study the formation of exciton-polariton condensates in two different lowest energy states at an exciton-polariton microcavity and the transition from the higher energy state to the ground state under pulsed and continuous wave excitation conditions by using various pump profiles.

In the second part of the thesis, we study the valley selection rules for the optical transitions from impurity states to the conduction band in two-dimensional Dirac materials, taking a monolayer of MoS$_2$ as an example, we focus on the weak light-matter coupling regime. We find the spectrum of the light absorption coefficients and calculate the photon-drag electric current density due to the impurity-band transitions.

\end{abstract}

% \includepdf[pages={1}]{add_pages/7.pdf}

\begin{romanpages}          % start roman page numbering
\tableofcontents            % generate and include a table of contents
\listoffigures              % generate and include a list of figures
\end{romanpages}            % end roman page numbering

\baselineskip=18pt plus1pt
\linespread{1.2}
 \setlength{\parskip}{1em}
 
%now include the files of latex for each of the chapters etc
\chapter{Introduction}\label{chap:intro}

\section{Excitons in semiconductors}

\subsection{Excitons in bulk semiconductors}
Light absorption in semiconductors results in the creation of a free electron in a conduction band and a hole in a valence band. A quasiparticle called excitons can be formed by the Coloumb interaction between a conduction-band electron and a valence-band hole, resulting in a bound state similar to a hydrogen atom. The energy spectrum of a hydrogen atom is a sum of the kinetic energy and binding energy of a hydrogen atom~\cite{moskalenko2000bose}, 
\begin{equation}
E_{H}=
\frac{\hbar^2k^2}{2(m_0 + m_p)}
-\frac{\hbar^2}{2m_0a^2_0n^2},    
\end{equation}
where $n$ is the principal quantum number, $m_0$ is the mass of a free electron and $m_p$ is the mass of a proton.
$a_0=4\pi\hbar^2\epsilon_0/(e^2m_0$) is the Bohr radius for a hydrogen atom. $\epsilon_0$ is the vacuum permittivity. The energy levels for the hydrogen atom can be used to define the binding energy for an exciton by replacing $a_{0}$ into $a_{ex}=4\pi\hbar^2\epsilon_0\epsilon_r/e^2m_{ex}$ 
, which is the Bohr radius for an exciton with the relative permittivity $\epsilon_r$ and the effective mass  $m_{ex}=\frac{m_em_h}{m_e+m_h}$, and $m_e(m_h)$ is the mass of an electron(a hole) in a conduction(valence) band and $\epsilon_0$ is the dielectric constant of the semiconductor. Such a replacement gives the energy of the $n^{th}$ exciton as
\begin{equation}
E_{ex}=
\frac{\hbar^2k^2}{2(m_e + m_h)}
-\frac{\hbar^2}{2m_{ex}a_{ex}^2n^2}
,
\label{bulk_exciton_energy}
\end{equation}
The substitution of the Coloumb attraction implies that the binding energy of the exciton is weaker than the binding energy of a hydrogen atom because the effective mass is significantly reduced and the permittivity for the semiconductor is larger than the permittivity in Vacuum. For example, in a GaAs semiconductor, the relative permittivity is the value of $\epsilon_r = 13.1$ and the effective mass is $m_{ex} = 0.0591$, resulting in the binding energy $\frac{m_{ex}}{m_0}\frac{13.6 \text{eV}}{\epsilon_r^2} = 4$ meV which is several order of magnitude smaller than the binding energy of the hydrogen atom $13.6$ eV. Due to the weaker binding energy of the excitons than the hydrogen atom, the Bohr radius of the exciton becomes much larger than the hydrogen Bohr radius. Based on the value of the Bohr radius, excitons can be categorized into two types, Frenkel excitons, and Wannier excitons. For the Frenkel excitons, the Bohr radius is less than the distance between the lattice sites, so that the electron and the hole are located in the same lattice point. Due to the large  Bohr radius for the Wannier excitons, the electron, and hole are separated by multiple unit cells.
% https://www.nextnano.com/manual/nextnano3_tutorials/exciton_1D.html

\subsection{Excitons in quantum wells}
\begin{figure}[ht]
\centering
\includegraphics[scale=0.6]{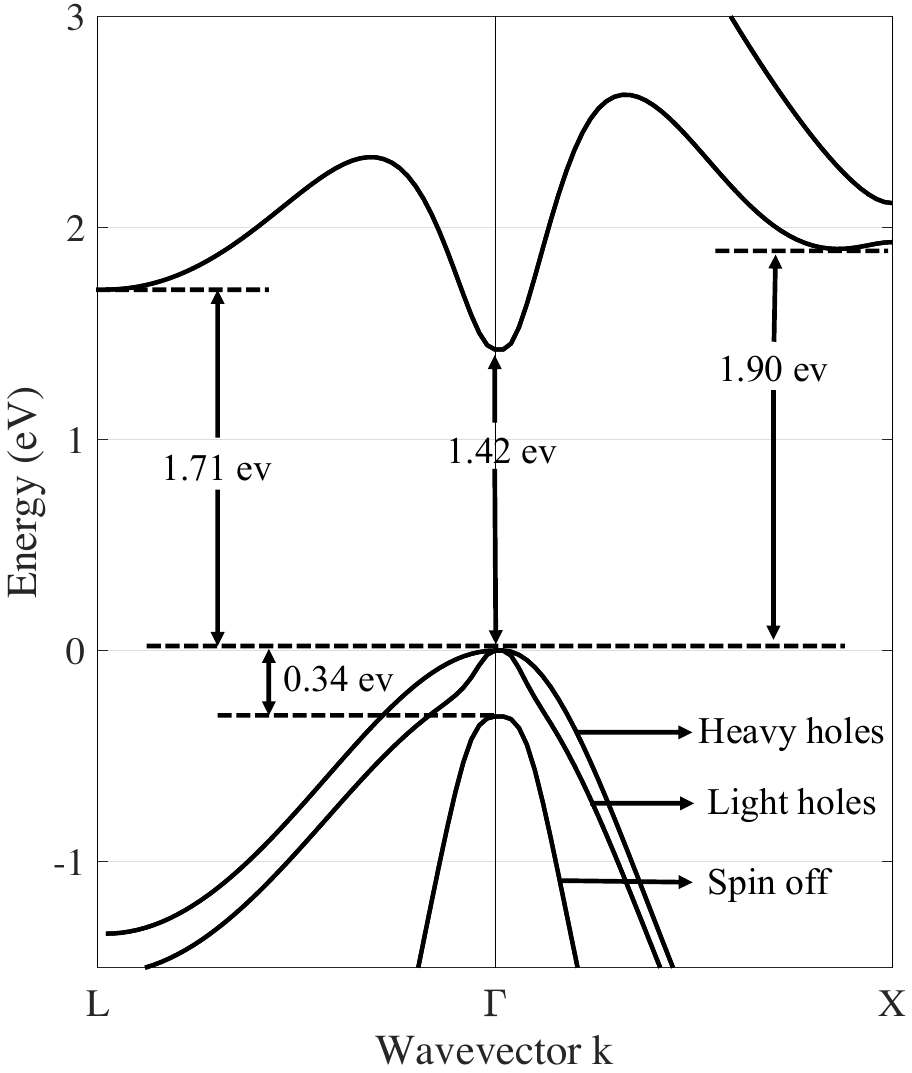}
\caption
[Band structure of GaAs]
{
Band structure of GaAs. The size of the direct band gap of GaAs at room temperature is 1.42eV. The valence band is split into subbands by light holes, heavy holes, and a split-off band.
}
\label{GaAs}
\end{figure}

\begin{figure}[ht]
\centering
\includegraphics[scale=0.7]{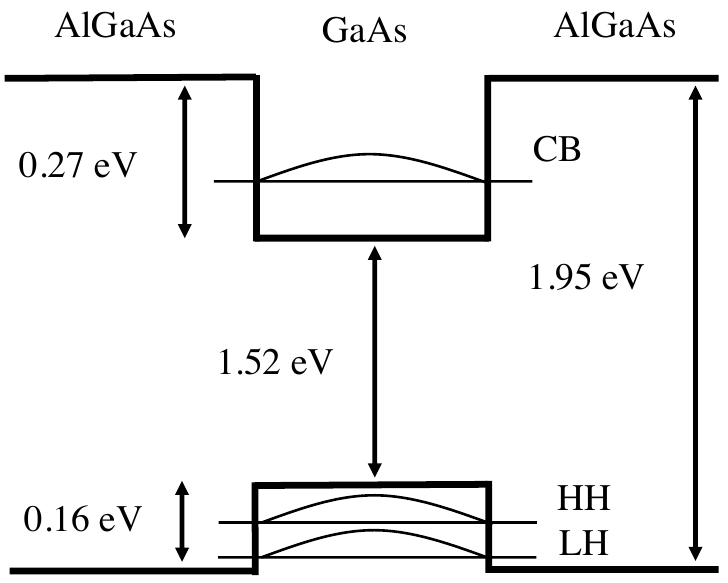}
\caption
[Structure of GaAs quantum well with AlGaAs barriers]
{
Energy band diagram of GaAs quantum well embedded between AlGaAs.
The band gaps for GaAs and AlGaAs at $0$K are 1.52eV and 1.95eV, respectively. The light hole(LH), heavy(HH), and the ground state of the conduction band(CB) are represented by their wave functions.
}
\label{QW}
\end{figure}

A quantum well(QW) is a thin layer of semiconductor, where the movement of particles is restricted in the direction perpendicular to the surface of the layer, whereas the particles freely move in the other directions. 
A QW usually represents a layer of a small band gap semiconductor embedded in between two other layers made of materials with higher band gaps, thus acting as barriers. 
These structures are usually fabricated with metal-organic chemical vapor deposition \cite{BHARDWAJ2022200} or molecular beam epitaxy \cite{dingle1974quantum}. A typical example of a quantum well is the gallium arsenide(GaAs) quantum well  sandwiched between the aluminum gallium arsenide(AlGaAs). GaAs is a direct band gap semiconductor in which the minimum of the conduction band lies at the same point in the k-space as the maximum of the valence band. The electronic structure of GaAs at $300$K is pictured in Fig.(\ref{GaAs}). The conduction band is separated from the valence band by $E_g = 1.519$eV at the temperature T$ = 0$K and $1.424 $eV at T$ = 300$K for GaAs. The valence band is split into the heavy hole band, the light hole band, and the split-off hole band, originating from the mixture of p-symmetry atomic orbitals. Fig.(\ref{QW}) shows the band structure for GaAs quantum well with AlGaAs barriers at $0$K. The thickness of the quantum well determines the confinement of the electron and hole wave functions between the barriers. Due to the confinement effect in GaAs quantum well, the bands for the light and heavy holes are split at $k=0$ as compared to the bulk GaAs semiconductor. Quantum confinement also affects the binding energy of excitons. The exciton binding energy in the quantum well case is increased by a factor of four compared to the case of bulk excitons \cite{haug2009quantum}.

\subsection{Excitons in 2D transition metal dichalcogenides}
\subsubsection{Transition metal dichalcogenides}
\begin{figure}
    \centering
    \includegraphics[scale=0.8]{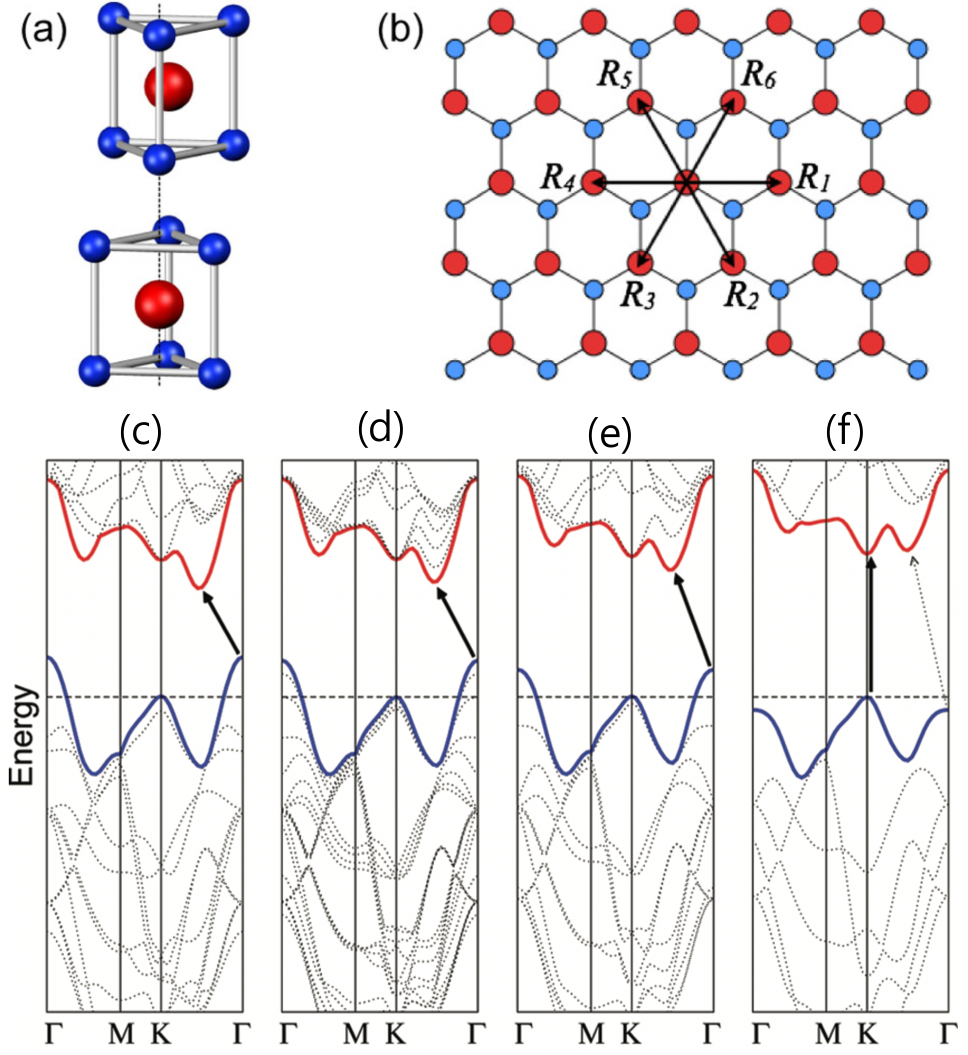}
    \caption
    [Band structure of MoS$_2$]
    {(a). The bulk unit cell of MoS$_2$. (b). Schematic of MoS$_2$ monolayer structure. Electronic band structures of the (c) bulk, (d) quadrilayer, (e) bilayer, and (f) monolayer MoS$_2$. The indirect band gap increases with the decrease in the number of layers, but
    the energy for direct excitonic transition rarely changes with the thickness of layers. This picture is taken from ~\cite{doi:10.1021/nl903868w}}.
    \label{BS_mos2}
\end{figure}

Transition metal dichalcogenides (TMDs) are a family of 2D materials with the form MX$_2$ where M corresponds to the transition metal elements (e.g. Mo, W) and X denotes the chalcogen elements (e.g. S, Se, Te). For the monolayer of TMD, such as MoS$_2$, the metal element molybdenum is located at the center of the trigonal prismatic geometry coordinated by six chalcogen atoms of sulfurs with strong covalent bonds. Multiple layers can be attached by weak van der Waals interaction. The structure of MoS$_2$ is depicted in Fig.(\ref{BS_mos2}). The band structure of MoS$_2$ monolayer displays a direct band gap located at the K points in reciprocal space, whereas bulk MoS$_2$ exhibits the maximum of the valence band located at the $\Gamma$ point and the minimum of the conduction band is located at a point between $\Gamma$, and $K$ points in the Brillouin zone, which indicates that bulk MoS$_2$ has an indirect band gap \cite{doi:10.1021/nl903868w}.

\subsubsection{Screening effect}
\begin{figure}[ht]
\centering
\includegraphics[scale=0.5]{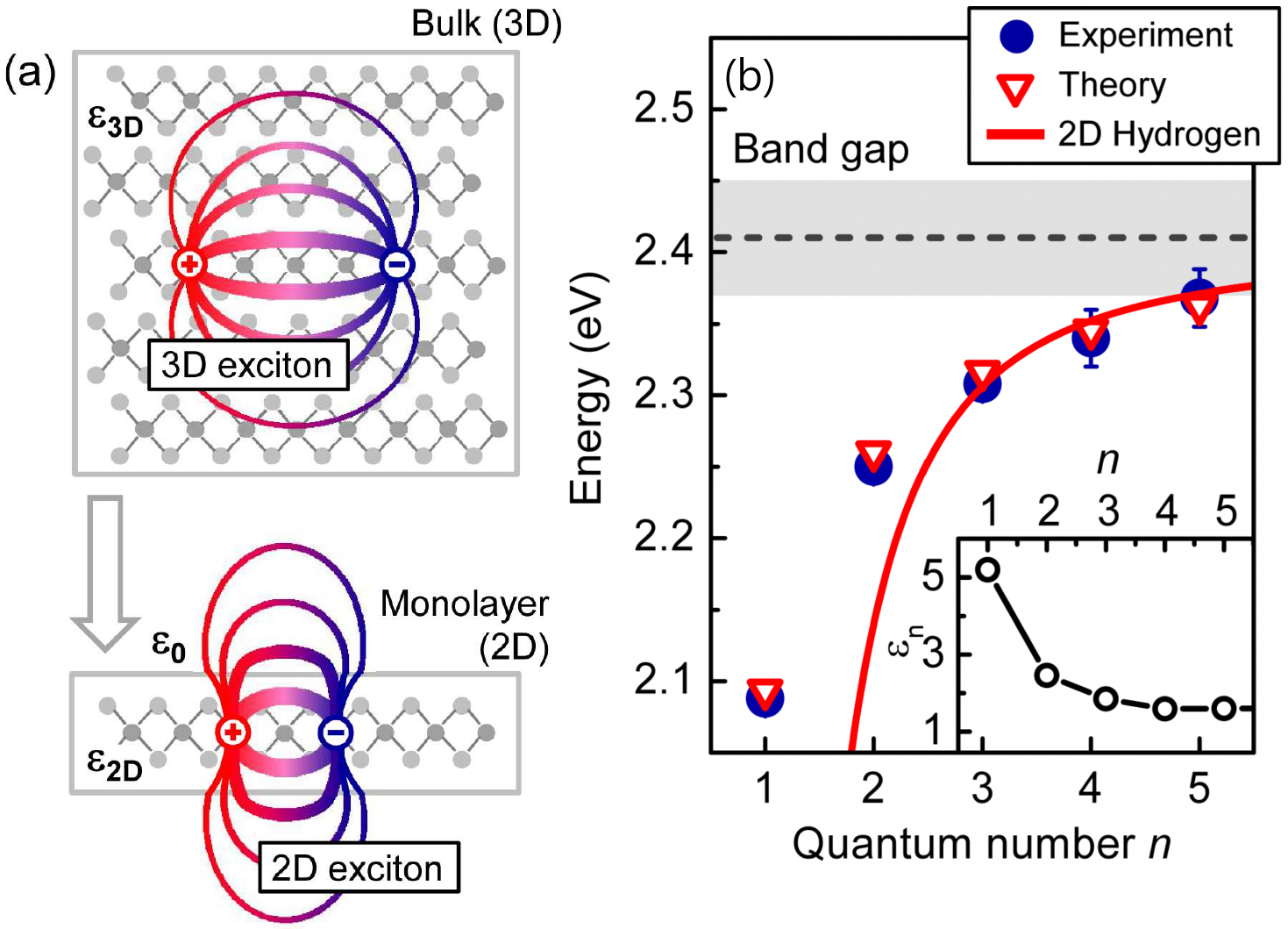}
\caption
[Screening effects in monolayer TMDs]
{
(a). Schematic representation of the exciton dielectric screening in the bulk and monolayer TMDs. (b). The energy of the states of the excitons in WS$_2$ monolayer as a function of the quantum number n. These pictures are adopted from~\cite{chernikov2014exciton}.
}
\label{screen}
\end{figure}

In bulk semiconductors, excitons can be described by the Bohr model of the hydrogen atom as negatively and positively charged particles bound to each other. However, to describe the energy of excitons in monolayer TMDs, the model for the hydrogen atom is not suitable anymore due to the change of the dielectric environment, resulting in the dielectric screening effects, which is depicted in Fig.\ref{screen}(a). An alternative model accounting for the inhomogeneous dielectric environment is introduced to demonstrate the exciton energies in the thin layers.
Another model can be built from the equation derived by Keldysh~\cite{keldysh1979coulomb} to describe the screened Coulomb potential between two point charges in a thin two-dimensional dielectric layer,
\begin{equation}
V(r) = 
-\frac{\pi e^2}{2r_0}
\left[
H_0(\frac{r}{r_0})
-Y_0(\frac{r}{r_0})
\right]
,
\end{equation}
where $e$ is the elementary charge, $r_0=d\epsilon/(\epsilon_a + \epsilon_b)$ is the effective screening length, $d$ is the thickness of the thin layer, $\epsilon$ is the dielectric constant for the thin layer, $\epsilon_a$ and $\epsilon_b$ are the dielectric constants for the surrounding medium and $H_0$ and $Y_0$ are the Struve and Neumann functions, respectively. Fig.\ref{screen}.(b) shows the experimental results for the exciton energies in WS$_2$, which agree greatly with theoretical calculations based on the Keldysh model. The ground state of the excitons shows the largest deviation compared to the hydrogen model for the Wannier excitons, and for the 2S state, it still shows a significant disagreement. In contrast, the higher excited states for the excitons fit the tendency of the 2D hydrogen model. This can be understood by the change in the spatial separation between an electron and a hole consisting of an exciton in a monolayer. As the quantum number of the exciton state increases, the exciton binding energy decreases, which leads to an increase in the distance between the electron and hole. This results in a larger fraction of the electric field leaking out of the layer, where the screening effect is strongly suppressed. Therefore, the hydrogen atom model can describe the $n=3-5$ excitons where the screening effect is highly suppressed, whereas the $n=1,2$ excitons exhibit non-hydrogenic behavior.

\subsubsection{Valley excitons}
The spin-orbit coupling in MoS$_2$ originating from d orbitals split the valence and conduction bands on subbands. The valence band splits extensively, whereas the conduction band remains almost spin degenerate. The splitting gives rise to two types of excitonic transition: \emph{A} exciton(ground state exciton) and \emph{B} exciton(higher spin-orbit split state). The energy of \emph{A} exciton is 100 to 200 meV lower than \emph{B} exciton energy, corresponding to the splitting of the valence band due to strong orbital coupling in MoS$_2$.  
% \textcolor{red}
% { The splitting by the spin-orbital coupling also leads to two types of excitons: bright and dark excitons. The electron-hole pairs are called bright excitons if the electron and hole have opposite spins, and they can recombine and emit a photon by conservation laws. In contrast, the dark excitons are formed by the pair of an electron and a hole with the same spins, and they cannot recombine through a photon emission because of the lack of momentum conservation. 
% }
%%%%%%%%%%%%%%%%%%%%%%%%%%%%%%%%%%%%%%%%%%%%%%%%%%%%%%%%%%%%%%%%%%%%%%%%%%%%%%%%%%%%%%%%%%%%%%%%%%%

\begin{figure}[ht]
\centering
\includegraphics[scale=0.5]{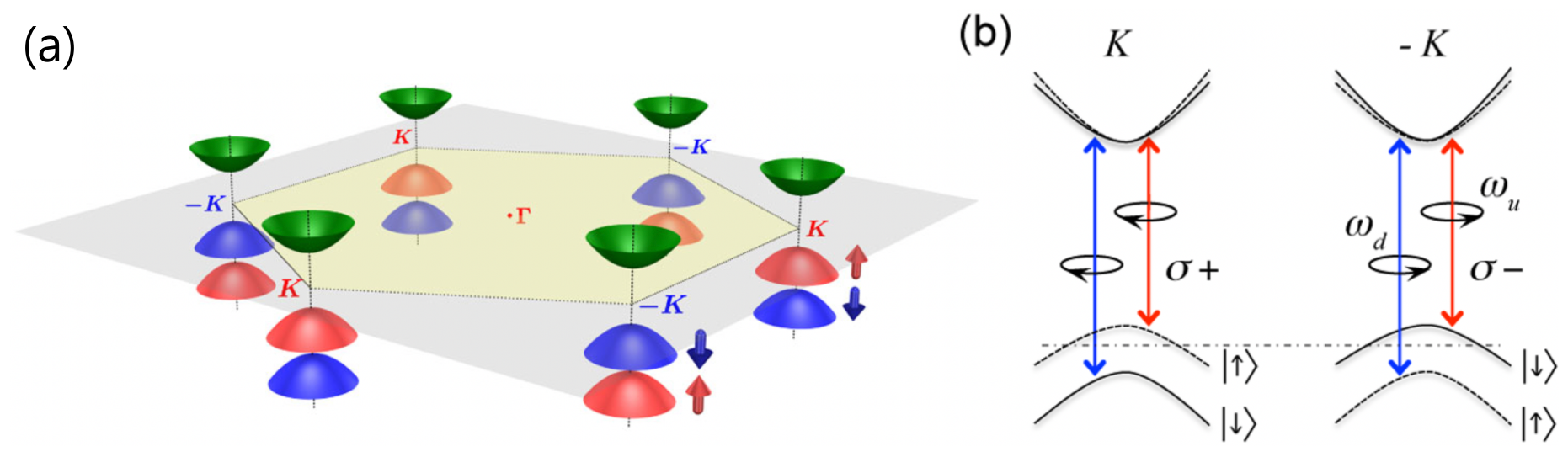}
\caption
[Band structures of a monolayer MoS$_2$]
{
(a). Schematic band structure of MoS$_2$ monolayer at K and -K points (b). Valley optical selection rules. The interband transitions at K(-K) points are exclusively coupled with photons with optical helicity $\sigma^{+}(\sigma^{-})$. Solid(Dahsed) lines denotes 
These pictures are adopted from~\cite{PhysRevLett.108.196802}.
}
\label{mos2_inversion}
\end{figure}

Another peculiar property of monolayer TMDs is the appearance of the valley degrees of freedom~\cite{PhysRevB.103.L161301}. The additional degree of freedom results from the broken inversion symmetry, leading to the nonequivalence of the two valleys at K and K' points of the two-dimensional hexagonal Brillouin zone as depicted in Fig.\ref{mos2_inversion}.(a). The broken inversion symmetry and the strong spin-orbital coupling lead to the valley-dependent optical selection rules, as shown in Fig.\ref{mos2_inversion}.(b). A right-handed polarized light can selectively excite the excitons in K valley. Conversely, a left-handed polarized light can excite the excitons in K' valley by satisfying the conservation of the angular momentum.

\section{Microcavity photons}
\subsection{Semiconductor microcavities}
In a semiconductor microcavity, the confinement of photons is attained by an optical resonator constructed by two opposing mirrors.
The microcavity consists of two distributed Bragg reflectors (DBRs) separated by the cavity layers. DBRs represent a one-dimensional structure of an alternating sequence of layers of semiconductors or dielectrics with different refractive indices $n_i$. The thickness $d_i$ of the layers is chosen by the relation $n_id_i=\lambda/4$, such that they fulfill the Bragg condition resulting in the constructive interference of the waves reflected from the multiple interfaces. In addition, the distance between the two Bragg mirrors should be proportional to an integer number of half-wavelength of light 
in the medium. By fulfilling the conditions, photons can be trapped in the cavity between the mirrors. 

\subsection{Dispersion of cavity photons}
The dispersion of photons in the planar microcavity is different as compared to the photon dispersion in free space due to confinement. In the case that photons are trapped in the z-direction, the dispersion relation can be written as
%%%%%%%%%%%%%%%%%%%%%%%%%%%%%%%%%%%%
\begin{equation}
E_{ca}=\frac{\hbar c}{n_{ca}}|\mathbf{k_{ca}}|=\frac{\hbar c}{n_{ca}}\sqrt{k_z^2 + k_\parallel^2},
\end{equation}
%%%%%%%%%%%%%%%%%%%%%%%%%%%%%%%%%%%%
where $\hbar$ is the reduced Planck constant, $c$ is the speed of light in the medium with reflective index $n_{ca}$, $k_z$ is the wave vector of light perpendicular to the cavity, and $k_{\parallel}^2 = k_x^2 + k_y^2$ is the in-plane components of the wave vector $\mathbf{k}_{ca}$. The component of the wave vector $k_{z}$, which is perpendicular to the cavity, is significantly larger than the in-plane wave vector $k_{\parallel}$, and the energy dispersion of cavity photon can be approximated with the Talyor expansion as
%%%%%%%%%%%%%%%%%%%%%%%%%%%%%%%%%%%%
\begin{equation}
E_{ca} \approx
\frac{\hbar c}{n_{ca}} k_\perp ( 1 + \frac{k_\parallel^2}{2k_\perp^2} )
=
E_{ca}(k_\parallel=0)+\frac{\hbar^2k_\parallel^2}{2m_{ca}},
\end{equation}
where $E_{ca}(k_{\perp}=0)=\hbar k_\perp c/n_{ca}$ and $m_{ca}=\hbar k_\perp n_{ca}/c$ is the effective mass of cavity photons .

% https://www.sciencedirect.com/topics/physics-and-astronomy/gallium-arsenides

\section{Exciton-polaritons}
Exciton-polaritons (polaritons) are quasiparticles created due to the strong coupling between cavity photons and quantum well excitons. Since polaritons are composite particles, polaritons inherit properties from their excitonic and photonic components, such as low effective mass and strong nonlinearity. As interacting bosons with low effective mass, polaritons are promising for investigating quantum collective phenomena, such as non-equilibrium Bose-Einstein condensation \cite{kasprzak06}, and the formation of quantum vortices at room temperatures \cite{lagoudakis2008quantized}.

\subsection{Dispersion of polariton states}
\begin{figure}[!ht]
\centering
\includegraphics[scale=0.7]{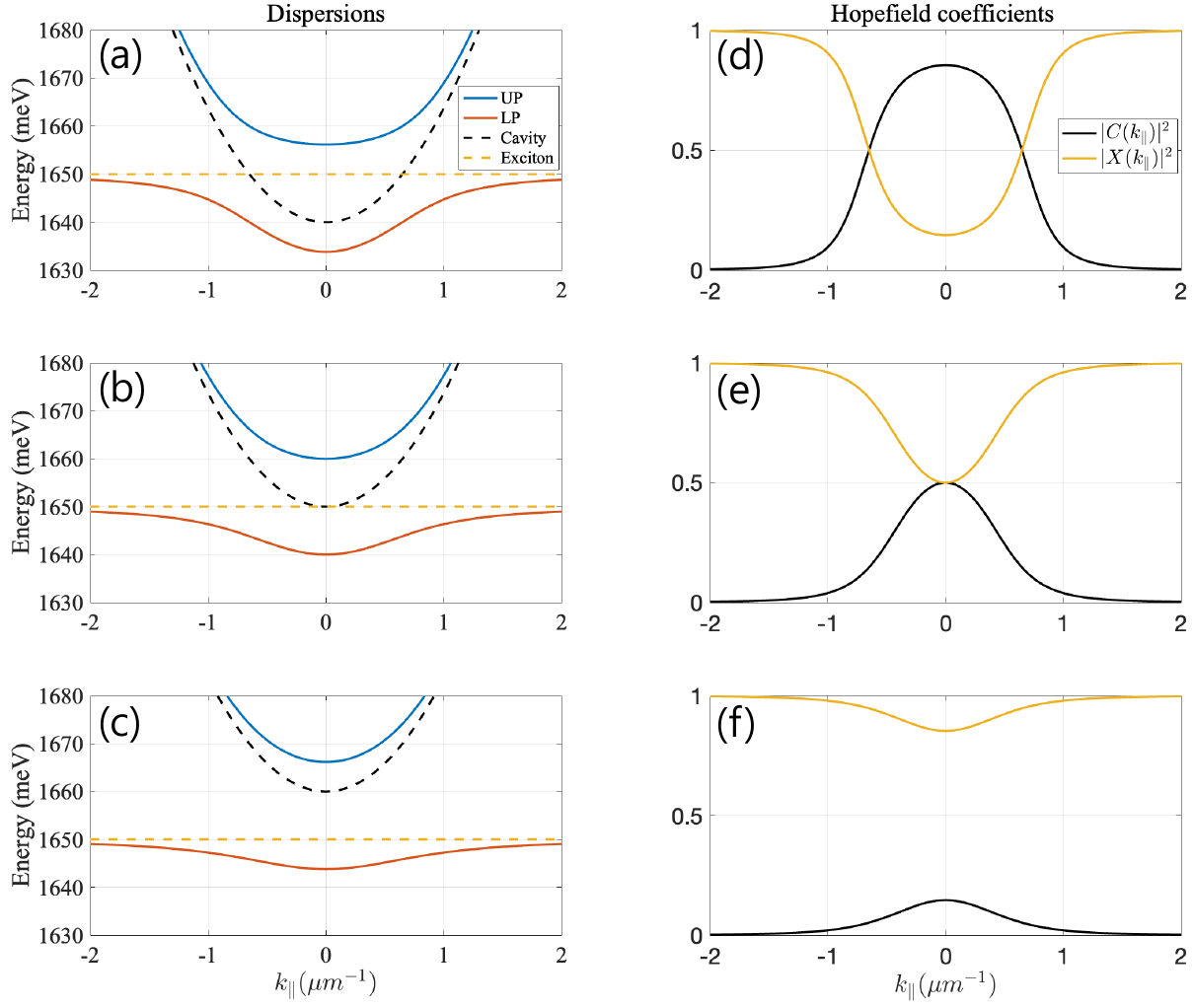}
\caption
[Polariton dispersions and Hopfield coefficients with various values of detuning]
{
(a-c). Energies of the upper polariton, lower polariton, bare cavity photon, and bare exciton as a function of in-plane wave vector $k_{\parallel}$ with different values of detuning -10 meV, 0 meV, and +10 meV, respectively. (d-e). The photonic and excitonic fractions of the Hopfield coefficients correspond to the dispersions.
}
\label{hope}
\end{figure}

The formation of polaritons can be described by the model of two coupled oscillators.
The Hamiltonian for the coupling reads
\begin{equation}
H=\sum_{k_\parallel}
[
E_{exc}(k_{\parallel})
\rho^{\dagger}_{k_{\parallel}}
\rho_{k_{\parallel}}
+
E_{cav}(k_{\parallel})
\phi^{\dagger}_{k_{\parallel}}
\phi_{k_{\parallel}}
+\hbar\Omega
(\rho^{\dagger}_{k_{\parallel}}
\phi_{k_{\parallel}}
+
\phi^{\dagger}_{k_{\parallel}}
\rho_{k_{\parallel}}
)
], 
\label{polariton_H}
\end{equation}
where $E_{exc}$ is the energy of an exciton in a quantum well, and $E_{cav}$ is the energy of a single cavity photon. The operators 
$
\rho^{\dagger}_{k_{\parallel}}
$ 
and 
$
\phi^{\dagger}_{k_{\parallel}}
$
are creation operators for excitons and cavity photons with the in-plane wave vector $k_\parallel$, and
$
\rho_{k_{\parallel}}
$ 
and 
$
\phi_{k_{\parallel}}
$
are the annihilation operators of the exciton and cavity photon, and $\hbar\Omega$ describes the interaction strength. The Hamiltonian can be also written in the matrix form,
\begin{equation}
H 
=
\begin{pmatrix}
E_{cav}&\hbar\Omega\\
\hbar\Omega&E_{exc}
\end{pmatrix}
\end{equation}
Two eigenvalues can be found by diagonalization of this matrix, yielding
\begin{equation}
E_{UP}(k_{\parallel})   
=
\frac{1}{2}
(E_{cav}(k_{\parallel}) +
E_{exc}(k_{\parallel}) 
)
+
\frac{1}{2}
\sqrt{
4(\hbar\Omega)^2 + (E_{cav}(k_\parallel)  - E_{exc}(k_{\parallel})
)^2
},
\end{equation}
\begin{equation}
E_{LP}(k_{\parallel})   
=
\frac{1}{2}
(E_{cav}(k_{\parallel}) +
E_{exc}(k_{\parallel}) 
)
-
\frac{1}{2}
\sqrt{
4(\hbar\Omega)^2 + (E_{cav}(k_\parallel)  - E_{exc}(k_{\parallel})
)^2
}.
\end{equation}
Here, $E_{UP}$ is the upper polariton state energy, which has higher energy than the lower polariton state energy $E_{LP}$. Both upper and lower polariton branches can be characterized by their photonic and excitonic fractions, $\abs{C(k_\parallel)}^2$ and $\abs{X(k_\parallel)}^2$, which are known as the Hopfield coefficients~\cite{deng2010exciton} and $\abs{C(k_\parallel)}^2+\abs{X(k_\parallel)}^2=1$. These coefficients obey
% These polaritons are part-light, part-matter quasiparticles with a composition that is given by the Hopfield coefficients
\begin{equation}
\abs{C(k_\parallel)}^2
=
\frac{1}{2}
(1-\frac{\Delta(k_\parallel)}
{\sqrt{\Delta^2(k_\parallel)+(2\hbar\Omega)^2}}
)
\end{equation}
\begin{equation}
\abs{X(k_\parallel)}^2
=
\frac{1}{2}
(1+\frac{\Delta(k_\parallel)}
{\sqrt{\Delta^2(k_\parallel)+(2\hbar\Omega)^2}}
),
\end{equation}
where $\Delta(k_\parallel)$ is the energy difference between cavity photons and excitons. Fig.(\ref{hope}) shows the polariton dispersions and the Hopfield coefficients with the different values of detuning. 

Due to their photonic properties, polaritons can rapidly propagate. Moreover, they can be optically excited and have short lifetime of the order of 10ps compared to the lifetime of excitons, which is on the order of 100ps up to a few ns. Photons can exchange their energy with optical phonons through the conservation of momentum and energy. In contrast, photons can not interact with acoustic phonons because the conservation laws are not satisfied. However, due to the excitonic component, polaritons can exchange their energy with acoustic phonons, resulting in the relaxation process in the lower polariton branch. Thanks to the polariton-polariton interaction, polaritons can accumulate at $k=0$ forming Bose-Einstein condensation.

\subsection{Polariton condensation}
\subsubsection{The basics of conventional Bose-Einstein condensation}
Bose-Einstein condensate(BEC) is a phenomenon characterized by the macroscopic occupation of the ground state in thermal equilibrium. Einstein extended the theory developed by Bose who studied the statistics of photons~\cite{bose1924plancks}. Thus Einstein describes a non-interacting gas of massive atoms in an equilibrium state and trapped in a box. He concluded that the lowest energy state 
is occupied by a massive number of particles
below some critical temperatures~\cite{einstein2005quantentheorie}. However, the interaction between particles and peculiar excitation spectra are essential to understand BEC. Bogoliubov developed a theoretical formulation for a weakly interacting Bose gas and predicted the spectrum behaves as phonon in low momentum region.~\cite{bogoliubov1947theory, ko2023bogolon}
The first experimental observation of BEC was performed in the gas of rubidium atoms cooled down to $170$nK by the group of Carl Wieman and Eric Cornell~\cite{doi:10.1126/science.269.5221.198}, which was followed by the observation of the BEC in other atomic gases: Lithium~\cite{bradley1995evidence} and Sodium~\cite{davis1995bose}.

\subsubsection{Bose-Einstein quasi condensation of exciton-polaritons}
Imamo\u{g}lu first suggested the theoretical description of the quasi-condensation of polaritons while studying the coherence of polariton Bose-Einstein condensation ~\cite{PhysRevA.53.4250}. The first attempts to observe the polariton condensation were demonstrated with a GaAs microcavity containing a low number of quantum wells. The dependence of lower polariton emissions on the pump intensity and time evolution of lower polariton emission at $k_{\parallel}=0$ ~\cite{doi:10.1126/science.1074464}.
The condensation of exciton-polaritons was clearly observed for the first time by Kasprzak et al.~\cite{kasprzak06}. In the experiment, they used a CdTe/cdMgTe microcavity containing sixteen quantum wells and the Rabi splitting was 26 meV. A polariton condensate was also observed in a planar GaAs microcavity in 2009 ~\cite{wertz2009spontaneous}. However, it is essential to point out that, unlike the conventional BEC, the polariton BEC is not an equilibrium state. It can even consist of several condensates emitting at different energies~\cite{krizhanovskii2009coexisting}.

\subsubsection{Excitaion of polariton condensates by resonant and nonresonant pumping}
\begin{figure}[!ht]
\centering
\includegraphics[scale=0.6]{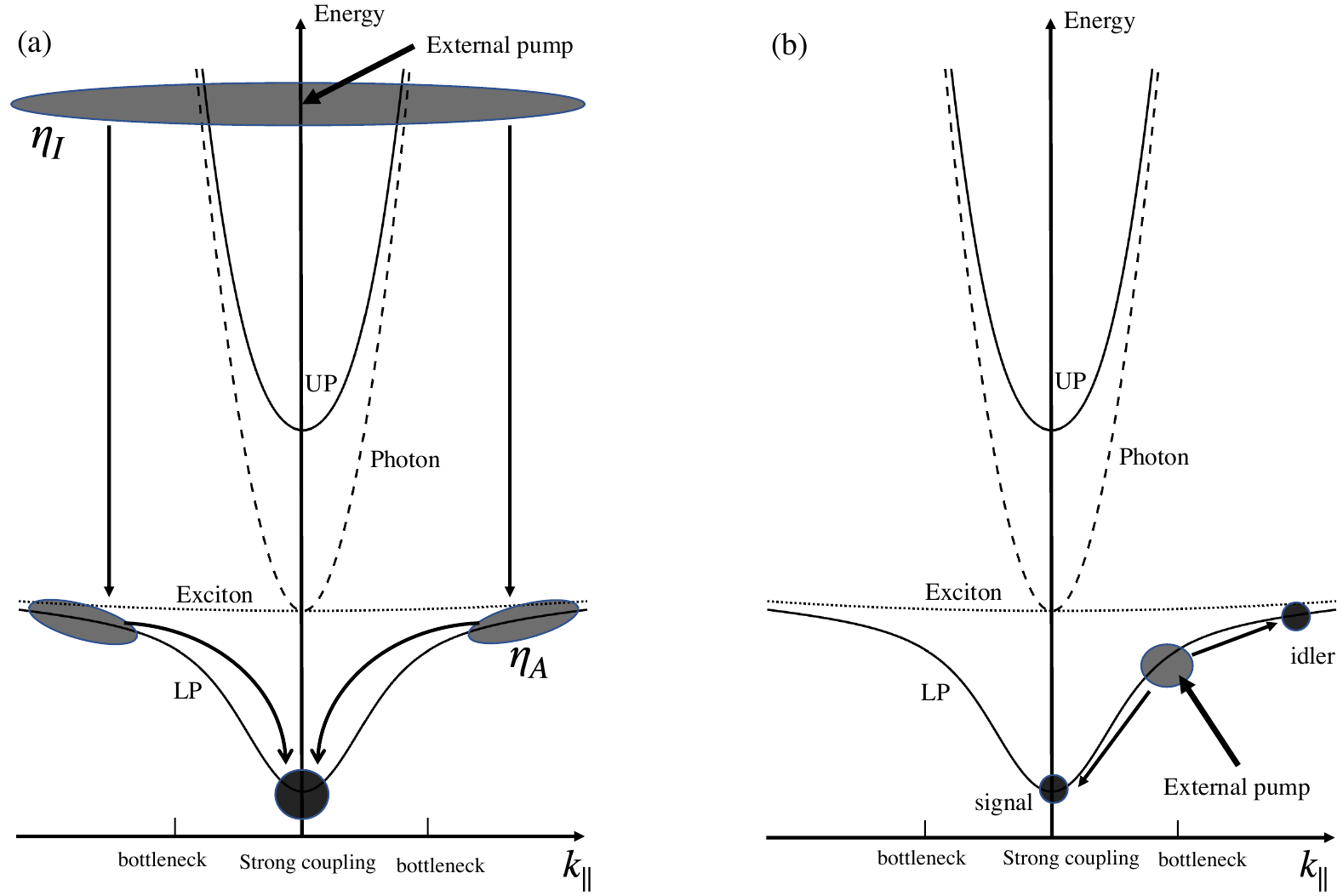}
\caption
[Resonant and nonresonant excitation]
{
(a). Schematic representation of the formation of the polariton condensate by nonresonant excitations. UP(LP) stands for the upper(lower) polariton branch, and $\eta_{I}(\eta_{A})$ represents the inactive(active) reservoirs. (b). Schematic of the resonant excitation.
}
\label{pumping}
\end{figure}

The first step to achieve the condensation of polaritons is to increase the polariton population, which can be done by external pumping. There exist two principal nonresonant and resonant pumping. In the nonresonant laser excitation scheme depicted in Fig.\ref{pumping}(a), free charge carriers are excited at the energy above the upper polariton branch, which then relax to the lower polariton branch via acoustic and optical phonon interaction~\cite{deng2010exciton}. %%%%%%%%%%%%%%%%%%%%%%%%%%%%%%%%%%%%%%
At high in-plane wave vectors, the polaritons for the lower polariton branch are mostly exciton-like, resulting in relatively long lifetimes, a large density of states, and heavy effective masses.
%%%%%%%%%%%%%%%%%%%%%%%%%%%%%%%%%%%%%%
%%%%%%%%%%%%%%%%%%%%%%%%%%%%%%%%%%%%%%
However, when the particles arrive at the dispersion inflection point where the actual strong coupling part of the lower polariton branch begins, polaritons become more photonic-like. Consequently, the lifetime of polaritons decreases by about two orders of magnitude, and the effective mass of polaritons is highly reduced by several orders of magnitude, resulting in a decrease in the density of states.
%%%%%%%%%%%%%%%%%%%%%%%%%%%%%%%%%%%%%%
From the inflection point, to reach the ground state at $k_{\parallel}=0$, polaritons should lose their energy about 5 to 10 meV. However, the relaxation to the ground state can not be attained by the interaction with the acoustic phonons since the required energy is larger than the energy of the acoustic phonons, which is usually less than 1meV. The acoustic phonon-mediated relaxation process takes longer than the polariton lifetime at the strong coupling regime, resulting in the particles accumulating at the inflection point. This phenomenon is referred to as the bottleneck effect, which was reported by Tassone~\cite{tassone1997bottleneck}. This obstacle can be suppressed by increasing the number of polariton pumps, resulting in more efficient energy relaxation to the ground state via polariton-polariton scattering above a certain
excitation density threshold.

The formation and time evolution of polariton condensates by the nonresonant excitation can be analyzed in detail based on a driven-dissipative Gross-Piaevskii equation coupled with reserovirs~\cite{lagoudakis2011probing},
\begin{equation}
\mathrm{i}\hbar\frac{\partial\psi}{\partial t}=-\frac{\hbar^2\nabla^2}{2m_p}\psi-\frac{\mathrm{i}\hbar}{2\tau_{p}}\psi+
(g + 
\frac{i\hbar R}{2}
)
n_A
\psi
+gn_{I}\psi
+\alpha|\psi|^{2}\psi,
\label{main}
\end{equation}
%%%%%%%%%%%%%%
\begin{equation}
\frac{\partial n_{I}}{\partial t}=P-\frac{n_I}{\gamma_I}-\frac{n_{I}}{\tau_{R}}
,
\label{ireservoir}
\end{equation}
%%%%%%%%%%%%%%
\begin{equation}
\frac{\partial n_{A}}{\partial t}=-\frac{n_A}{\gamma_A}+\frac{n_{A}}{\tau_{R}}-R|\psi|^{2}n_{A}\label{areservoir}    
\end{equation}
Eq.(\ref{main}) describes the polariton condensate,
$\psi$ is the coherent polariton field in the lower polariton branch.
$m_p$ is the mass of polaritons, $\tau_p$ is the lifetime of polaritons, and $g$ is the energy exchange term between the reservoir particles and the polariton condensate. $R$ describes the rate of exciton scattering into the polariton condensate, and $\alpha$ is nonlinear repulsive polariton interactions. 

Eq.(\ref{ireservoir}) describes the rate equation for the inactive reservoirs. Indeed, the external pump creates an inactive exciton. By the nonradiative recombination, the inactive excitons disappear with the decay rate $1/\gamma_I$ or they can be replenished into the active reservoir with $1/\tau_R$. The rate equation for the active reservoir is given by Eq.(\ref{areservoir}). The particles in the active reservoir disappear radiatively at the rate $1/\gamma_A$, or they are scattered down to the polariton condensate at the rate $R$.

Another method to create a polariton condensate is using a resonant excitation into the lower polariton band~\cite{PhysRevB.62.R16247}. The resonant pump results in optical parametric oscillation, which can help polaritons overcome the bottleneck effect. Optical parametric oscillators are resonators based on an optical gain through a nonlinear medium. The nonlinear resonator generates a second-order nonlinear process that a coherent pump passing by the medium, produces the coherent signal and idler outputs~\cite{Byer1977}. 
Polaritons can experience a stimulated scattering process when injected by a coherent pump on the lower polariton branch with the angles near the inflection points depicted in Fig.~\ref{pumping}. (b).
%%%%%%%%%%%%%%%%%
By exciting the polaritons with a resonant pump above the threshold at the momentum $k_{p}$ of the inflection point, two polaritons can scatter into two different states, which are a zero momentum and a higher momentum $2k_{p}$ in such a way that the polariton can achieve the energy and momentum conservation.
%%%%%%%%%%%%%%%%%
Due to the bosonic statistic of the polaritons, a stimulated scattering process takes place, and the interaction between polaritons allows the particles to be populated into the ground state at $k=0$. 
{
% It is peculiar to note that, in contrast to non-resonant excitation, the phase and polarisation of the excitation laser is imprinted on the resonantly excited polariton population.
}
%%%%%%%%%%%%%%%%%
% This scattering process is much different than under nonresonant excitation, where the population of excitons and polaritons with high momentum is significant, since only a small fraction of polaritons can reach the minimum of the lower polartion branch. 
% }
% \textcolor{blue}
% {%%%%%%%%%%%%%%%%%
% It has been experimentally shown that phase of the pump polartions do not influence the phases of the signal nor idler polartions, 
% hence the created condensate is independent from the pump and only condition that $\phi_{idler}+\phi_{signal} = 2\phi_{pump}$ has to be fulfilled. 
% %%%%%%%%%%%%%%%%%
% }

\chapter{Partial revivals of localized condensates in distorted lattices}\label{chap:condensates}

\section{Introduction}
Through classical trajectories, coherent states~\cite{perelomovbook86} of simple quantum systems show nondestructive dynamics, but the wave packets are generally dissipative in time. This spreading reverses if the energies of the system are rational numbers, leading to complete or fractional revivals of the quantum states as, e.g., in the case of the Brown states~\cite{brown73} of the hydrogen atom~\cite{PhysRevA.42.6308}. 
Particles loaded in a flat band (FB) of a periodic potential~\cite{Derzhko:2015aa}. exhibits unique quantum transport properties. The particles form the compact localized state (CLS)~\cite{Sutherland:1986aa} by occupying only a few neighboring sites of the lattice in an ideal case described by the tight-binding model. The CLS is an eigenstate of the single-particle Hamiltonian, and thus it does not propagate nor does it spread in time. However, in reality, perfect localization is hard to achieve due to the approximate flatness of the band, interparticle interactions, and finite lifetime of particles.
These relentless obstacles lead us to the natural questions: Are FBs actually feasible? How do reasonably small deformations of a FB affect the dynamics of initially perfect CLSs? Moreover, can one at all speak about any (quasi-)CLSs in realistic circumstances?

We address these issues by studying the dynamics of exciton-polaritons (later polaritons) loaded in a nearly perfect FB~\cite{ko2020partial}. 
Under external coherent or incoherent excitation, polaritons might condense~\cite{kasprzak06, balili07}, demonstrating macroscopic occupation of a particular state---phenomenon also referred to as polariton lasing. 
The state in question can be localized or trapped, and there exist multiple experimental scenarios of polariton trapping, either in a single state or in a periodic network. In particular, polaritons can be exposed to spatially-periodic acoustic waves~\cite{cerdamendez10} or created in etched microcavities. 
%cerdamendez13
Polariton condensation in artificial periodic potentials has recently become a remarkably active field of research. 
Polaritons develop new transport properties if loaded in honeycomb~\cite{jacqmin14}, kagome~\cite{masumoto12,gulevich16}, or 1D and 2D Lieb~\cite{baboux16, klembt17, whittaker18} lattices, occasionally forming topologically protected~\cite{karzig15,nalitov15,bardyn15,stjean17,chunyanli18, sun2019exciton} and single-particle quasi-flat bands (which we will also refer to as FBs in what follows). 
%\alexei{AA: Most examples of topological flatbands are only quasi FB, and I feel we need to specify that.}
%

Several questions remain open and disputable in this field. In particular, polariton condensates in FBs possess a relatively short coherence length. It is unclear if this is the consequence of a disorder or if condensate fragmentation is a generic property of out-of-equilibrium systems loaded in a FB.
However, a periodic long-range order can appear spontaneously in resonantly-driven cavities~\cite{gavrilov18}.
Moreover, condensed out-of-equilibrium particles should not necessarily occupy the lowest energy state (typically the $\Gamma$ point). There emerge $\pi$-condensates (at the edge of a band) in 1D potentials~\cite{lai07} and $d$-condensates in 2D square lattices~\cite{kim11}. The choice of the condensate phase is controlled by the polariton-polariton interaction, which can also lead to a space-time intermittency regime in microcavities with periodic potentials (lattices)~\cite{yoon19}.

\section{Tight-binding model on kagome lattice}

\begin{figure}[b!]
\centering
\includegraphics[width=0.6\textwidth]{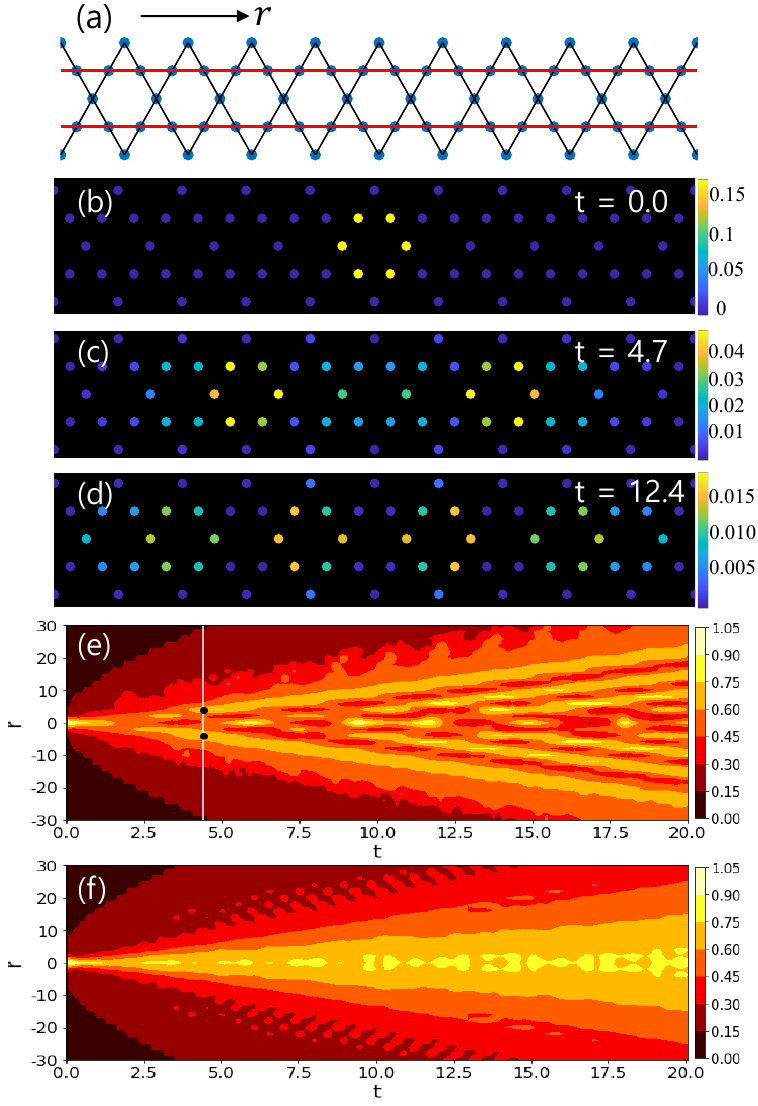}
\caption
[Tight-binding description for kagome chains]
{Tight-binding description. (a) \textcolor{black}{A kagome chain, with red (horizontal) bonds having half the hopping integrals of the other (black) bonds.}
State of the system: (b) $t=0$ (arb. units), the initial state is the exact CLS; (c) $t=4.7$; (d). $t=12.4$. (e-f)  Density plot of CLS revivals evolution in time with a kagome chain and kagome lattice, respectively: every point of the y-axis is a hexagon. The color denotes the value of the overlap with a CLS. The vertical line in (e) indicates the time shown in (c), and the black dots represent the two copies of the original CLS. }
\label{fig11}
\end{figure}

We start by using a regular tight-binding model on a kagome chain. 
It has a FB in the case of uniform hoppings. By adjusting the hopping coefficients in horizontal bonds [see Fig.~\ref{fig11}(a)], we lift the FB degeneracy and study the unitary evolution of the system initialized in one of the CLSs, which occupies sites of a single hexagon [Fig.~\ref{fig11}(b)].
\textcolor{black}{Namely, the hoppings on the horizontal bonds, indicated by the red color in Fig.~\ref{fig11}(a), have half the value of the non-horizontal bonds.
This mimics the lattice distortion of the continuous model discussed below. The period of the revivals and the spreading velocity is mostly affected by altering the ratio of the hopping coefficients.}

%We modify the hopping parameters for dilated neighbouring bonds, shown in  Fig.~\ref{fig11}(a).} 
%\textcolor{green}{YR: All red bond are dilated? I do not understand Fig.1(a).}
 
%

Due to the distortion, the CLS ceases to be an exact eigenstate, and it spreads over the chain with time. 
We observe a signature of revivals of the CLS at short times [Fig.~\ref{fig11}(c)] and nonperfect twins of the original CLS that appear at shifted positions. 
\textcolor{black}{We note that this is a partial revival since the whole state of the quantum system does not fully coincide with its original state, and it is characterized by the presence of a few CLS condensates further away from the creation center.}
Eventually, the CLS is completely destroyed [Fig.~\ref{fig11}(d)]. 
This destruction occurs faster for stronger distortions.

Figures~\ref{fig11}(e) and~\ref{fig11}(f) illustrate the entire process of the CLS partial revivals and their eventual destruction in a kagome chain (e) when we leave a single layer of hexagons composed of triangles and full 2D kagome lattice (f).
We plot the overlap of every hexagon in the chain with a CLS and the time evolution of the overlaps.
One observes relatively clear signatures of revivals at short times, but they are gradually suppressed at longer times.
Similar but weaker signatures also occur in a 2D kagome lattice, where a stripe of sites with different couplings is introduced. 
The revivals of the initial CLS in the tight-binding model are weaker than the ones seen in the continuous model, which we discuss next.

\section{The formation and propagation of polaritons in the lattice}
\begin{figure}[b!]
\centering
\includegraphics[width=0.6\textwidth]{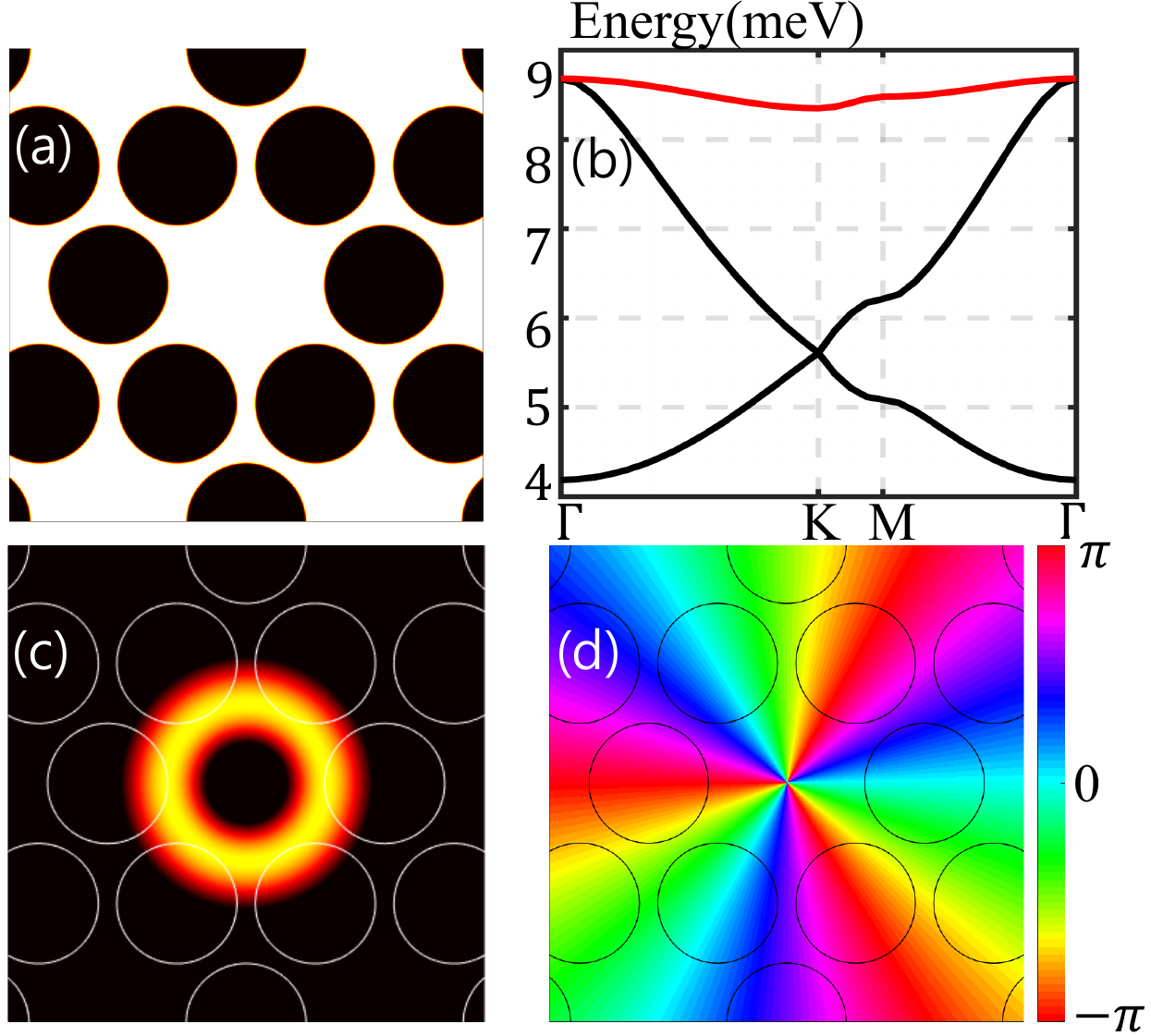}
\caption
[System schematic]
{System schematic. (a) A compact localized state can be created at the center of the 2D Kagome lattice, occupying six pillars. 
(b) The band structure shows the two lowest dispersive bands and the highest non-dispersive ones. 
(c) The intensity of Laguerre-Gaussian (LG) pumping that aligns well with the hexagonal structure of the kagome lattice. 
(d). Phase of the LG beam.}
\label{fig2}
\end{figure}

We study the formation and propagation of polaritons coupled with an incoherent excitonic reservoir, using the evolution equations
\begin{eqnarray}
\mathrm{i}\hbar\frac{\partial\psi}{\partial t}=\mathcal{H}\psi+\mathcal{P}-\frac{\mathrm{i}\hbar}{2\tau_{p}}\psi+\frac{\mathrm{i}\hbar G n_{R}}{2}\psi+\alpha|\psi|^{2}\psi+\textcolor{black}{gn_{R}\psi},
\label{eq:pl}\\
\frac{\partial n_{R}}{\partial t}=P_{in}-\frac{n_{R}}{\tau_{R}}-G|\psi|^{2}n_{R}, 
\label{eq:rev}\\
\mathcal{H} = -\frac{\hbar^2}{2m_p}\nabla^2 +V
\end{eqnarray}
where $\psi$ is a polariton macroscopic wave function, $\mathcal{H}$ is the Hamiltonian, responsible for the propagation of the particles with the effective mass $m_p$ in a kagome lattice potential $V$, shown in Fig.~\ref{fig2}(a);
$\mathcal{P}$ is the coherent pumping term; $\tau_p$ and $\alpha$ are the polariton lifetime and polariton-polariton interaction constant, respectively;
\textcolor{black}{g is polariton-reservoir interaction constant;}
$n_R$ is the reservoir particle density, which is incoherently pumped by 
\textcolor{black}{the continuous wave (CW) and homogeneous in space} 
term $P_{in}$ and has a finite lifetime $\tau_R$; the reservoir and polaritons are coupled by a phenomenological constant $G$. 

Taking $V=0$ inside the lattice sites (pillars) and $V=30$ meV outside the pillars, we first solve the eigenvalue problem $\mathcal{H}\psi=E\psi$ in the framework of the continuous model, and we find the band structure, shown in Fig.~\ref{fig2}(b).
Unlike the tight-binding model, we note that the third band here is only approximately flat.
To create a CLS of the \textcolor{black}{nearly} flat band, we employ a coherent Laguerre-Gaussian (LG) pump at the center of the lattice, as shown in Fig.~\ref{fig2}(c), which is given by
\begin{equation}
    \mathcal{P}=P_{0}\left(\frac{r}{R}\right)^l L^{0}_{l}\left(\frac{r^2}{R^2}\right) exp\left[-\left(\frac{r^2}{R^2}\right)+\mathrm{i}(l\phi-\omega_{0}t)\right],
    \label{eq:LG}
\end{equation}
where $P_{0}$ is the pump amplitude, $r$ is the radial distance from the center of the plaquette, $\phi$ is the azimuth angle, $R=0.7\,\mu\mathrm{m}$ is the radius of the pumping ring, $l=3$ determines the phase difference between neighboring lattice sites (adjacent pillars), which in our case is $\pi$ [see Fig.~\ref{fig2}(d)]. 
Furthermore, $\omega_{0}$ is the pump frequency, which we put equal to the frequency of the third band of the kagome lattice at the $\Gamma$ point.
\textcolor{black}{For short pulses, the spreading in energy is greater than the width of the flat band, so that the exact coincidence of frequencies is not necessary for the CLS excitation.}

\textcolor{black}{We note that it is convenient to excite CLSs in the kagome lattice with the Laguerre-Gaussian pump, as compared to the Lieb lattice~\cite{PhysRevB.98.161204}, due to the absence of parasite pumping of the sites outside the CLS, i.e., the sites, where the population should be zero as a result of destructive interference.}

We apply this coherent pump for a short time, thus making a short pulse of about $0.5\,\mathrm{ps}$ duration, to generate the CLS and let it evolve with the support of the continuous background incoherent pumping $P_{in}$.
\textcolor{black}{
In the absence of $P_{in}, $ the population of the system quickly decreases. The subthreshold pumping $P_{in}<P_{th}=(G\tau_{R}\tau_{P})^{-1}$ helps to prolong the existence of the condensate by compensating the losses due to the finite lifetime.
Nevertheless, this background incoherent pump does not hinder the revival phenomenon, as shown below.
}

\textcolor{black}{It is essential that the main excitation comes from the resonant Laguerre-Gaussian pulse. We assume this pulse alone to be strong enough to create the substantial macroscopic occupation of CLS, such that its evolution can be described by mean-field equation even without additional non-resonant pumping. The non-resonant pumping improves the applicability of the Gross-Pitaevskii equation by additional feeding of this macroscopic state. It is assumed that the number of particles is still macroscopic after revivals, so the semiclassical description remains valid. 
On the other hand, it is necessary to keep the background non-resonant pumping below the threshold. Otherwise, the effect would be masked by the spontaneous condensate formation.}

\section{Distorted lattice}
\begin{figure}[!ht]
\centering
\includegraphics[width=0.70\textwidth]{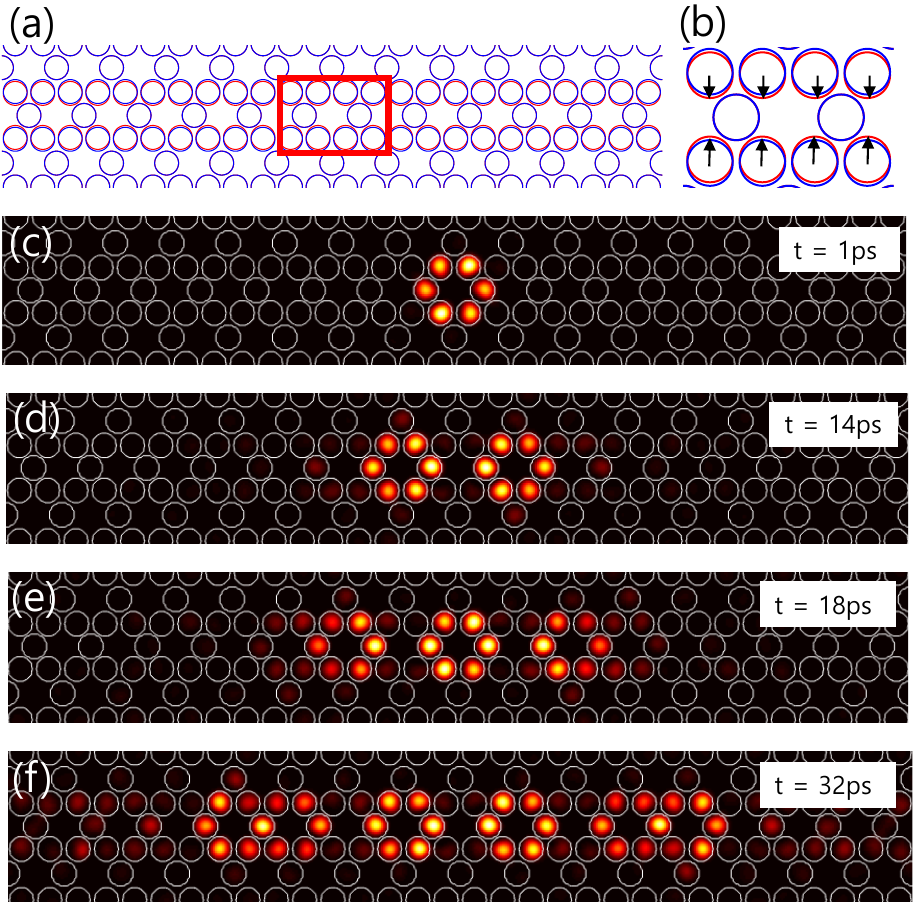}
\includegraphics[width=0.70\textwidth]{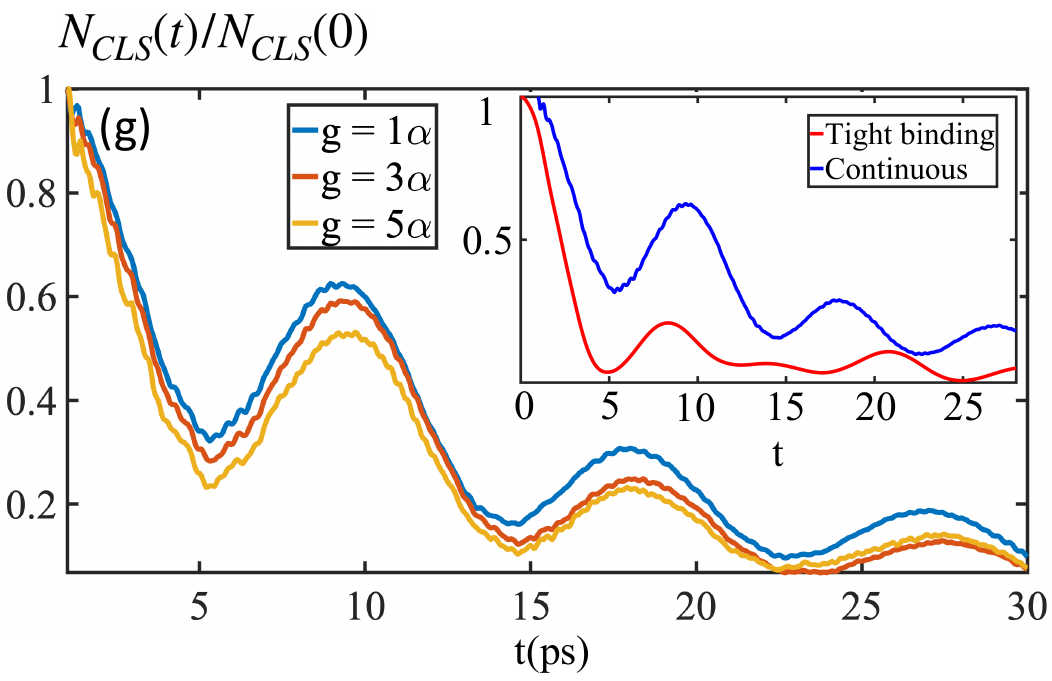}
\caption
[Evolution of a polariton condensates in a kagome strip]
{
(a) The kagome $y$-squeezed strip. (b) Red pillars are squeezed vertically. Snapshots of CLS at 1~ps (c), 14~ps (d), 18~ps (e), and 32~ps (f). 
% (Visualization 1). 
The vertical lines indicate times, shown in (d-f). (g) Temporal evolution of the number of particles in CLS at the center of the strip \textcolor{black}{for different polariton-reservoir interaction constants}: a manifestation of the partial revivals. 
(Inset) Comparison of the time evolution of the CLS weight in the tight-binding (red) and continuous (blue) model.
}
\label{fig4}
\end{figure}

Similarly to the tight-binding case, we introduce anisotropy in the system, as it is shown in Fig.~\ref{fig4}, and solve Eqs.~(\ref{eq:pl}) and (\ref{eq:rev}) numerically. 
As a result, we observe (i) quasi-1D propagation of the CLSs in specific directions, which can be seen as partial jumps, and (ii) revivals of the CLSs at their original position and different positions in space. 

%
%
% \clearpage

We study the kagome strips squeezed in $y$-direction, as shown in Figs.~\ref{fig4}(a,b). $x$ squeezed kagome strips is disscused in appendix ~\ref{cls_appendix}. 
%\textcolor{black}{(the results of the $x$-squeezing are discussed in Appendix)}.
\textcolor{black}{We note that experimentally, two methods exist to squeeze the lattice. One way is to etch the lattice with a defective line, where the distances are slightly different. Another way is to apply unidirectional mechanical stress to the lattice, which will cause effective deformation of the lattice. 
}
%, and Fig.~\ref{fig4}(b)}blue pillars stand for the non-deformed kagome lattice, while the red pillars are shifted vertically.

In this work, we simulate the former method. \textcolor{black}{The lattice constant is 3$\mu\mathrm{m}$. The radius of the pillars is $0.65\mu\mathrm{m}$ and the pillars are shifted as $0.1\mu\mathrm{m}$.}
The revival of the CLS on a kagome $y$-squeezed strip is shown in Fig.~\ref{fig4}(c-f). The LG beam shines in the center of the lattice for $0.5$ ps. Figure~\ref{fig4}(c) shows the particle density when the coherent pump is switched off. 
The condensate now propagates both to the left and the right, and the CLS at the center of the strip completely disappears at about $14\,\mathrm{ps}$. 
At the same time there appear new CLSs next to the center. See Fig.~\ref{fig4}(d).
Further on, the condensate continues to propagate and creates new CLSs in neighboring sites, but the CLS at the center is partially restored at $32\,\mathrm{ps}$, see Fig.~\ref{fig4}(e).
Figure~\ref{fig4}(f) shows that the CLS at the center of the strip disappears again, and the condensate reaches the boundary of the system.
\textcolor{black}{Figure~\ref{fig4}(g) shows the influence of the interaction constant $g$ on revivals. 
By increasing $g$, the number of particles in the strip decreases.
This figure also proves that the revival phenomenon still occurs with higher polariton-reservoir interaction.}

We considered a clean system and disregarded any structural disorder. Typical potential disorder in polariton lattices of 0.2 meV is much smaller than the height of 30 meV of the potential barriers in the kagome lattice. Thus we do not expect strong suppression of the revivals.
We also note that the effect has an essentially local character: the particles created by the initial pulse spread and come back involving a few lattice sites only. Therefore, experimentally one can try to find a place in the lattice with the particularly weak disorder to observe the revivals.

\section{Summary of the chapter}
We reported on the peculiar unexpected dynamics of bosonic condensates loaded in a distorted kagome lattice, where the transport of particles manifests itself in jumps and partial quantum revivals of compact localized states. 
In the tight-binding model, the dynamics of the initial CLS take place by its reappearance in the neighboring sites, but the revival of the original CLSs is not observed. In contrast, in the continuous model, the deformation of the nearly flat band by squeezing the lattice in either a horizontal or vertical direction leads to visible revivals of the polariton wave packet. 
%The kagome $y$-squeezed strips show better revivals than the kagome $x$-squeezed strips, in addition, polaritons in $x$-squeezed strips propagate faster. 

\textcolor{black}{
The physical mechanism of the revival phenomenon is constructive interference of waves reflected from the lattice while propagating uni-dimensionally. 
In the usual one-dimensional systems, the back-scattering leads to the Anderson localization, resulting in disordered localized states of particles. In contrast, the localized states in FB lattices are prepared in specific CLS form, and the back-scattering produces revivals of these CLSs without change in their shape and size.}
%\textcolor{red}{The physical mechanism of the revival phenomenon is the proximity to a flatband and therefore the presence of \emph{fast} propagating dispersive modes and \emph{slow} CLS modes, that are responsible for the revivals, and that both get excited during the propagation of the original excitation.}
A background pump can also support this effect, opening new possibilities for developing all-optical logical elements based on polariton condensates in nearly flat bands.

\textcolor{black}{We expect that our work will stimulate the research towards building new quantum devices, such as kagome optomechanical
lattice~\cite{Wan:17}. Moreover, exciton-polariton delocalization and revivals can be used for information transfer and storage~\cite{cite-key}.}
\chapter{Observation of a quantized vortex vanishment in exciton-polariton system}\label{chap:vortex}

\section{Introduction}
A superfluid behaves differently as compared to conventional fluids due to its macroscopic quantum coherence. To describe the superfluid state, the single-valuedness of the wave function is introduced and it gives rise to the quantized circulation in units of $\hbar/m$, where $\hbar$ is the reduced Planck constant, and $m$ is the mass of the particle. An integer multiple of $2\pi$ phase winding around the vortex core is carried in the superfluid with the quantization of circulation and the density of the superfluid is depleted. 

Starting from the first prediction of quantized circulation in superfluid helium by Onsager~\cite{onsager1949statistical}, the creation of quantized vortices has been extensively studied in conventional single-component bosonic superfluids such as liquid helium~\cite{yarmchuk1979observation,sachkou2019coherent} and cold atomic gases~\cite{matthews1999vortices,abo2001observation}. Due to the macroscopic phase winding, a quantized vortex is considered to be a topologically stable object. Therefore, these vortices can be potentially utilized as quantized information bits~\cite{choi2022observation}. The macroscopic phase winding associated with a quantized vortex makes it a topologically stable object; hence, these vortices can be potentially used as quantized information bits~\cite{choi2022observation}. Their topological stability has extensively drawn attention regarding particle injection, and decay dynamics~\cite{ruutu1997annihilation,adams1987vortex,bretin2003dynamics}. 

Exciton-polaritons (later polaritons) provide an excellent platform for investigating the stability of a topological vortex in a non-equilibrium superfluid~\cite{lagoudakis2008quantized,lagoudakis2009observation}. A polariton is a quasiparticle created as a result of the strong coupling between a cavity photon and a quantum-well exciton in a semiconductor microcavity. Polaritons inherit most of the properties from their constituents. Due to the photonic component, polaritons have a short lifetime which results in the non-equilibrium physics~\cite{szymanska2006nonequilibrium,wouters2007excitations}, and the excitonic constituent gives rise to the interaction between polaritons. Moreover, the bosonic nature of polaritons allows the formation of macroscopic quantum phenomena such as polariton condensation~\cite{kasprzak06}, quantized circulation~\cite{liu2015new}, and superfluidity~\cite{amo2009superfluidity}. In particular, numerous theoretical~\cite{liew2007excitation, ma2018vortex, yulin2016spontaneous, ostrovskaya2012dissipative} and experimental methods have been widely developed to generate quantized vortices in polariton systems, including the circular grating structures~\cite{hu2020direct}, resonant excitation by Laguerre–Gaussian (LG) beams~\cite{boulier2015vortex}, and optical parametric oscillations~\cite{sanvitto2010persistent}. 

In addition, polariton superfluids have been also investigated to achieve quantized vortex states even under nonresonantly pumped LG beam excitation~\cite{PhysRevLett.122.045302}. The combination of the photonic constituent of polariton and the non-equilibrium nature of the polariton condensates creates a unique opportunity for studying the time-resolved dynamics of these vortices. In fact, the creation of multiply charged vortices~\cite{alperin2021multiply,cookson2021geometric} and decay of vortices have been investigated, such as vortex–anti-vortex pair annihilation~\cite{cancellieri2014merging,hivet2014interaction}, vortex annihilation through spiraling out of condensates~\cite{wouters2010superfluidity}, and the destruction of the condensate itself~\cite{sanvitto2010persistent}. However, the dynamics associated with vanishing a single polariton vortex within the polariton condensate lifetime have not been reported.

\section{Formation of vortex states}
In this part, we show the vanishment of a single quantized vortex in a polariton superfluid. To create a vortex, we used a nonresonant LG beam, which imprints the orbital angular momentum (OAM) of the pump beam to the polariton condensate~\cite{PhysRevLett.122.045302}. The energy-integrated photoluminescence (PL) image shows that the vortex contains an incomplete depletion of the superfluid density at the center. Based on the careful spectral analysis of the vortex, we discovered that the polaritons form two different modes of condensation and the incomplete depletion of the superfluid density at the vortex core is an artifact of the time-averaged measurement of the multimode condensate, as evidenced by energy-resolved interferometry experiments. We found that only the higher energy state of the multimode condensate carries angular momentum, referred to as a quantum vortex, by separately extracting the phase profile of the wave functions in the two different states. Furthermore, time-resolved spectroscopy exhibits that the excited (vortex) state relaxes to the stationary ground state in a pulsed LG beam excitation, which serves as a direct experimental signature of a quantized vortex vanishment from the system. Such a transition is demonstrated by modifying the diameter of the LG pumping beam. 

We can attain a steady state where the excited- and ground states coexist under continuous-wave (CW) pumping. The experimental observations are supported by a theoretical model based on the driven-dissipative Gross-Pitaevskii equation coupled with incoherent and coherent reservoirs; hence, we suggest a mechanism for OAM transfer from the pump beam to the polariton condensate that is based on the microscopic coherent properties of the latter. Furthermore, we describe the transition of the system from the first excited state characterized by the vortex and a concrete OAM to the intermediate regime, when two lowest-energy states coexist, and finally, to the ground-state condensate with zero OAM after vanishment of the vortex.

%%%%%%%%%%%%%%%%%%%%%%%%%%%%%%%%%%%%%%%%%%%%%%%%%%%%%%%%%%%%%%%%%%%%%%%%%%%%%%%%%%%%%%%%%%%%%%%%%%%%%%%%%%%%%

\begin{figure}[b!]
\centering
\includegraphics[width=1\textwidth]{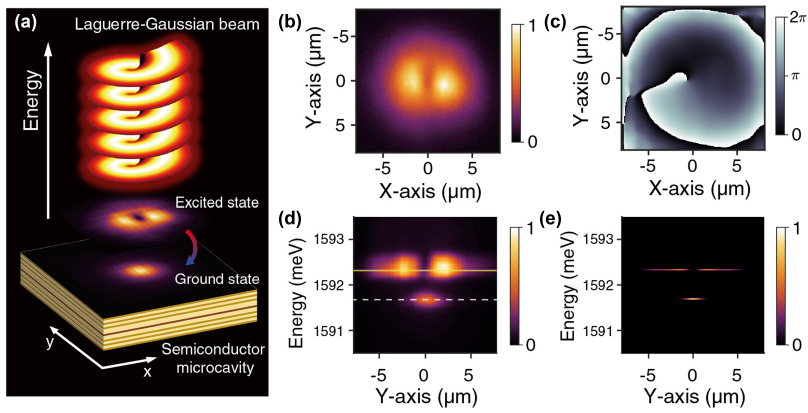}
\caption
[System schematic]
{
(a) Schematic of the system optically pumped with a nonresonant LG beam with OAM $\abs{l}=1$. Above the pumping threshold, polaritons are condensed, forming two energy states with different density distributions. (b) Energy-integrated PL image of the condensates above the pumping threshold. (c) Phase map of (b) shows $2\pi$ phase winding. (d) Real-space spectrum along line $x = 0$ in (b). (e) The calculated real-space spectrum for the condensate.
}
\label{FIG1}
\end{figure}
\begin{figure}[b!]
\centering
\includegraphics[width=1\textwidth]{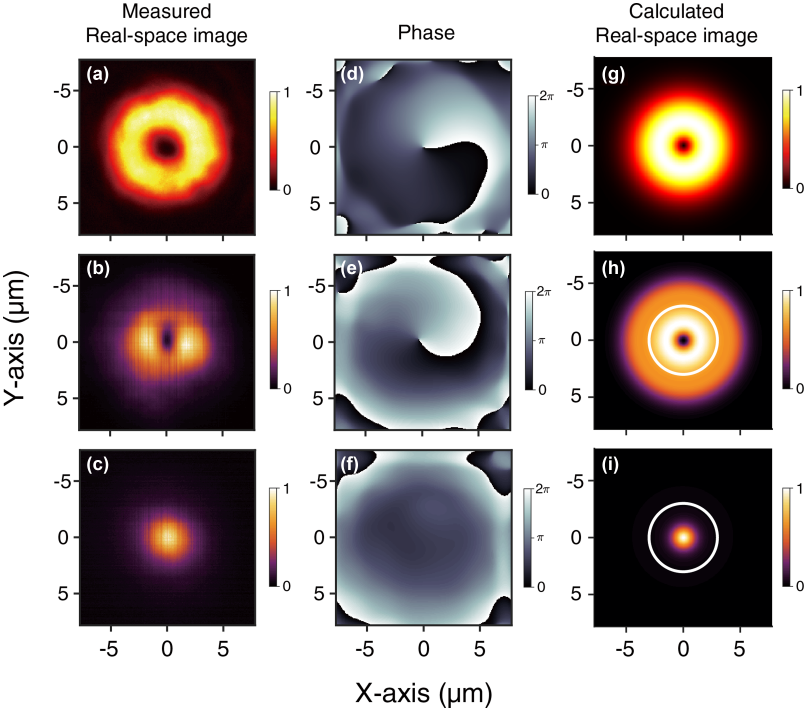}
\caption
[Pump and condensates profiles.]
{
Profiles of the pump and multimode condensates (a) LG pump intensity distribution; energy resolved PL images of (b) excited state at $1592.37meV$, and (c) ground state at $1591.69 meV$ under pulsed excitation; (d) phase map of LG beam having OAM $l = +1$; phase maps of (e) excited state and (f) ground state. A $2\pi$ phase winding in (e) indicates a quantized vortex in an excited state. Panels (g), (h), and (i) show calculated real-space images of LG beam excitation and polariton condensates when the system is stationary (either ground or excited state). White circles in panels (h) and (i) illustrate the nominal diameter (6.0 $\mu m$) of the LG beam corresponding to the maximum intensity spots of the simulated beam (g).
}
\label{FIG2}
\end{figure}

In the experiment, a microcavity that contains GaAs quantum wells(QW) is used and for the nonresonant excitation, the LG beam is illuminated to the microcavity, as depicted in Fig.~\ref{FIG1}(a). In the microcavity, an optical cavity is constructed between a pair of distributed Bragg reflectors with a set of GaAs/AlAs QW stacks. To reach as low as $5K$, the microcavity sample was put inside a cryostat(with an optical window). A Ti: sapphire laser which contains an energy of $1.72eV (720 nm)$ was used as a pulsed and a CW excitation to pump the electron-hole plasma into the QWs nonresonantly. An LG beam was generated as the Gaussian laser beam passed a $2\pi$ winding phase mask, which resulted in an OAM of the $l=+1$ beam Fig.~\ref{FIG2}(a). Polaritons formed condensates with a quantized vortex above the pumping threshold. The energy-integrated PL image is shown in Fig.~\ref{FIG1}(b) and the phase map is depicted in Fig.~\ref{FIG1}(c) under pulsed excitation. In the phase map, the measurement of $2\pi$ phase winding implies that the superfluid carries a circulation of $\hbar/m$. Fig.~\ref{FIG1}(d) shows the real-space spectrum along $x = 0$. From the spectrum, the condensate is not in a uniform static state but two distinctly quantized states. We will refer to the higher- and lower-energy states as the excited and ground states for simplicity. For the theoretical description of the experiment, we used a driven-dissipative Gross-Pitaevskii equation coupled with incoherent and coherent reservoirs, where the latter carries the same OAM as the LG beam. We assume that the OAM of the pumping source is almost erased in the scattering processes, and only a small amount remains. The OAM is revived in the polariton condensate owing to the interaction with a weak coherent reservoir and the coherent properties of the condensate. Fig.~\ref{FIG1}(e) exhibits a simulation result of multistate condensate similar to that in the experiment.

Energy-resolved spatial interferometry based on the tomographic measurement protocol~\cite{PhysRevB.82.073303, PhysRevB.82.045304, PhysRevB.80.121309, PhysRevX.5.011034, Buller_2016} is employed to study the phase information and density profile in each state. The experimental procedure was similar to the modified Mach-Zehnder interferometry, which is particularly utilized to visualize quantized vortices. Fig.~\ref{FIG1}(d) shows the excited and ground state which contains the energy $1592.37 meV$ and $1591.69 meV$ and they are represented with solid and dotted lines, respectively. To represent the spatial densities of polaritons in the excited- and ground energy levels, we show energy-resolved PL intensity under pulsed excitation which is depicted in Fig.~\ref{FIG2}(b) and Fig.~\ref{FIG2}(c). The excited state shows the density minimum in the center and the ground state reveals a density distribution in Gaussian form. Fig.~\ref{FIG2}(e) and Fig.~\ref{FIG2}(f) show the phase maps attained from the interference patterns of these two states. The excited state exhibits a $2\pi$ phase winding with respect to the density minimum point, indicating a quantized vortex. The phase winding of the vortex Fig.~\ref{FIG2}(e) was identical to the OAM of the LG pump Fig.~\ref{FIG2}(d). However, the phase map for the ground state indicated no such phase winding, and the phase was relatively flat aside from the radial gradient Fig.~\ref{FIG2}(f). In other words, the ground state did not carry any vortices. The simulated LG beam Fig.~\ref{FIG2}(g) resulted in polariton condensate formation at the two energy levels Fig.~\ref{FIG1}(e). The real-space emission profiles in the steady state for each of these energy levels are shown in Fig.~\ref{FIG2}(h) and Fig.~\ref{FIG2}(i), exhibiting consistency with the experiment. (To simulate the formation of the ground- and excited states, we used the CW excitation instead of the exact femtosecond-order pulsed excitation). 

\section{Pulsed-excitation regime}

\begin{figure}[b!]
\centering
\includegraphics[width=1\textwidth]{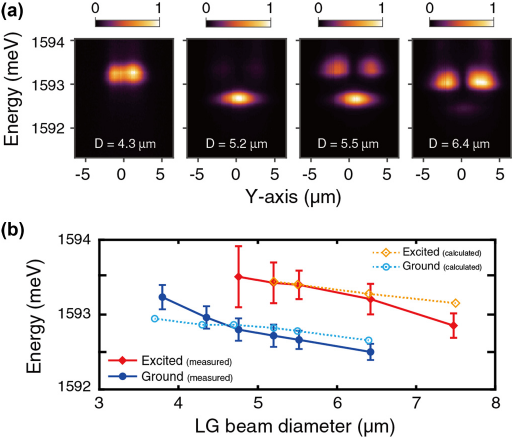}
\caption
[Energy of vortex and ground states with various beam diameters]
{
(a) Real-space spectra of PL profiles in the pulsed-excitation regime for different pumping beam diameters of $4.3, 5.2, 5.5$, and $6.4 \mu m$. Beam diameter is defined as the distance between two maxima of laser profile Fig.~\ref{FIG2}(a). (b) The energy of ground- (blue circle) and first excited (red diamond) states as a function of the beam diameter. The numerical simulation of ground- (open circle) and excited states (open diamond) are plotted.
}
\label{FIG3}
\end{figure}

The time-integrated real-space spectra measurement reveals that the relative intensity of the PL emission from the two states depends on the diameter of the LG beam, as shown in Fig.~\ref{FIG3}(a). Below the diameter of the beam $4.5\mu m$, only the ground state is populated. However, the population of the excited state starts to increase as the beam diameter increases, and, at a sufficiently large diameter, the ground state becomes almost invisible. 

This can be explained by a confinement effect: if the beam size is sufficiently small, the excited state is pushed out of the potential trap. This can be understood as a confinement effect: the excited state is pushed out of the potential trap if the beam spot is sufficiently small. The confinement effect is reduced when the diameter of the beam increases, resulting in a redshift Fig.~\ref{FIG3}(b). 

\begin{figure}[ht]
\centering
\includegraphics[width=1\textwidth]{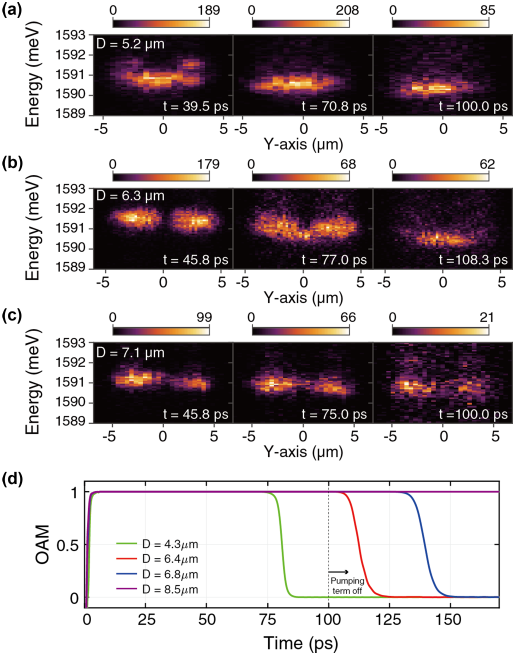}
\caption
[Evolution of polariton vortex with different pump diameter]
{
Time-resolved spectroscopy of polariton vortex in pulsed LG pump regime for different beam radii: (a) $5.2$, (b) $6.3$, and (c) $7.1 \mu m$. Data were measured along a straight line through the center of the beam (at $x = 0 \mu m$). (d) Simulation of OAM evolution for different beam diameters with the same initial condition. The pumping term was turned off after $100 ps$ to simulate the pulsed excitation case.
}
\label{FIG4}
\end{figure}

The condensate can be formed in the excited state when the energy of the first excited state is sufficiently low enough to fit inside the potential trap.  This behavior was confirmed via numerical calculations. The numerical calculations and experimental results are depicted in Fig.~\ref{FIG3}(b). The figure exhibits the energy of the excited state and ground state with different pump diameters. Due to the reduced blueshift with smaller confinement, the slope of the data downwards with an increase in the diameter of the beam. To determine the exact nature of the multistate condensate under the pulsed excitation, the time-integrated measurement is not suitable since it does not allow us to figure out where polaritons occupy the two distinct energy states sequentially or simultaneously. We performed time-resolved real-space spectroscopy which measures the relative intensity of the two different states in particular time intervals to distinguish between the two possible scenarios, depicted in Figs.~\ref{FIG4}(a) - Figs.~\ref{FIG4}(c). The results show that the excited state which carries the vortex firstly forms and then sequentially the ground state appears. Eventually, the system experienced a transition from the vortex carrying excited state to the ground state. Note that in a smaller LG beam diameter Fig.~\ref{FIG4}(a), polaritons occupy the “Gaussian” ground state dominantly (appearing as single lobe along the $y$ axis at $x = 0$) in the measured time evolution. However, in the case of intermediate LG beam diameter Fig.~\ref{FIG4}(b), polaritons occupy the “annular” excited state (appearing as double lobes along the $y$ axis at $x = 0$) initially, and they experience a transition to the ground state as time proceeds. In the case of a larger LG beam diameter Fig.~\ref{FIG4}(c), polaritons show only the excited state over the whole time duration. Simulations also confirmed these measurements. Our simulations indicate that the transition phenomena were accompanied by the competition between two energy states and the lowering of the potential barrier ~\cite{PhysRevB.92.035305} as time proceeds under the pulsed excitation. In fact, the transition to the lower-energy state occurred due to scattering processes ~\cite{PhysRevB.82.245315}. Polaritons try to occupy the vortex core, which is the center of the potential trap when the population of the ground state increases. Therefore, the excited state competes with the growing ground state to occupy the vortex core and the ground state to push the excited state out of the interior of the potential barrier. In pulsed excitation, the gradual depletion of the exciton reservoir over time diminishes the potential trap as compared to CW excitation. Then, the excited state is pushed out faster, and the system remains in the ground state. The entire process can take longer than the condensate lifetime, and polaritons decay before the transition occurs for sufficiently large beam diameters, shown in Figs.~\ref{FIG4}(c). For smaller beam diameters Fig.~\ref{FIG4}(a), before polaritons completely decay, the excited state can be pushed out more easily; hence, an earlier transition to the ground state can be observed. Although the total OAM of the polaritons will not reach complete unity in practice because of inherent fluctuations in the sample and the nonideality of the LG beam, the simulation explains well what we see in the experiments Fig.~\ref{FIG4}(d): a smaller beam size resulted in the sooner loss of the total OAM of the system. We investigated the dependence of the OAM on the size of the LG beam and its intensity in the regime of the pulsed exaction. 

\begin{figure}[b!]
\centering
\includegraphics[width=1\textwidth]{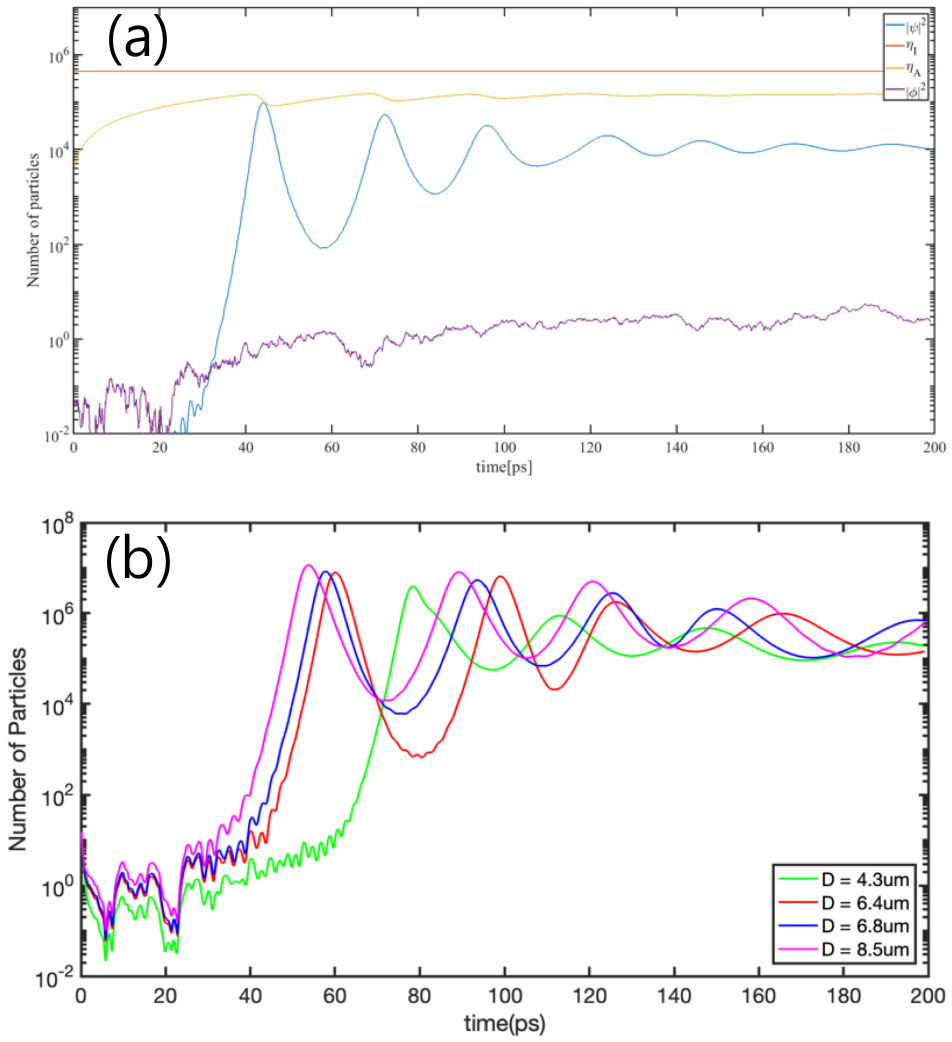}
\caption
[Number of particles in condensate and reservoirs]
{
(a). Number of condensate and reservoir particles in the system as functions of time for a single simulation. Blue line: number of particles in polariton condensate. Red line: number of particles in the inactive reservoir. Yellow line: number of particles in active reservoir. Purple line: number of particles in the coherent reservoir. Initial states were as follows: zero initial condition for the coherent and active reservoir, steady state for the inactive reservoir, and random initial condition for polariton condensate. (b). Number of particles in condensate for different pump diameters. A random initial condition was applied for polariton condensate; The initial conditions are the same as (a).
}
\label{FIG111}
\end{figure}

To numerically investigate the evolution of the exciton-polariton condensate, we used the driven-dissipative Gross–Pitaevskii equation as follows: 
\begin{equation}
i\hbar\partial_t\psi
=
-\frac{\hbar^2\psi^2}{2m}\psi
-
\frac{i\hbar}{2\tau}\psi
+
\frac{i\hbar R\eta_A}{2}\psi
+
\frac{i\hbar g}{2}\phi
+
\alpha\abs{\psi}^2\psi
+
\beta(\eta_A + 2\eta_I)\psi
+
\beta\abs{\phi}^2\psi.
\end{equation}
Here, $\psi(r,t)$ is the complex-valued macroscopic wave function of the lower-branch polaritons; the effective mass of the polaritons was $m=0.7\times10^{-4}m_e$, where $m_e$ represents the free electron mass, the particle lifetime $\tau = 5 ps$, and the repulsive polariton–polariton interaction was $\alpha=10^{-3}meV\cdot\mu m^2$. To include the energy relaxation and scattering dynamics of particles created by the off-resonant optical excitation, we considered two incoherent reservoirs: an active reservoir $\eta_A$ and an inactive reservoir $\eta_I$. The polariton condensate was replenished by the active reservoir controlled by a phenomenal parameter $R=0.02 ps^{-1}\cdot\mu m^2$. The intensity of the repulsive interaction between the condensate and reservoir particles was $\beta=10^{-2}meV\cdot \mu m^2$. Furthermore, we assumed a small fraction of coherent excitation $phi$ to be inherited by the condensate (with $g=1ps^{-1}$). This allowed one to achieve identical vorticity in the excitation laser and polariton condensate. The coherent polaritons interacted with the condensate, as enabled by the term $\beta\abs{\phi}^2\psi$ (which is small and negligible). The following equation governs the dynamics of the active reservoir:
\begin{equation}
\partial_t\eta_A=\frac{\eta_I}{\lambda}
-
R\abs{\psi}^2\eta_A
-
\frac{\eta_A}{\gamma}.
\end{equation}
Here, the lifetime of the reservoir particles is $\gamma=300 ps$. The active reservoir is refilled by the inactive reservoir at a rate of $\lambda^{-1}=120ps^{-1}$. 
The inactive reservoir is pumped directly using an external incoherent pumping source $\abs{P_i}$. We assumed that the lifetime of the inactive reservoir particles is the same as that of the active reservoir. Hence, the following equation is obtained:
\begin{equation}
\partial_t\eta_I=\abs{P_i}-\frac{\eta_I}{\gamma}-\frac{\eta_I}{\lambda}.    
\end{equation}
The weak coherent excitation field obeys the evolution equation as follows:
\begin{equation}
i\hbar\partial_t\phi
=
iP_c exp[-i\omega(t)t]
-
\frac{i\hbar R_{in}}{2}\abs{\phi}^2\phi.
\end{equation}
Here, $P_c$ is the coherent pumping source, which breaks the central symmetry and hence transfers the winding number information from the laser pump to the condensate. The frequency of this coherent pump $\omega(t)$ was selected randomly at each time step according to the uniform 
distribution in range [ min($E_{pl}/\hbar$), max($E_{pl}/\hbar$) ], where $E_{pl}$ is the spectrum of the lower polariton branch. To saturate the gain of this coherent reservoir, we introduced a nonlinear loss term controlled by the parameter $R_{in} = 1 ps^{-1} \mu m^2$ (last term in the equation above). The LG pumping term used in the simulation (both coherent pump $P_c$ and incoherent pump $P_i$  ) reads 
\begin{equation}
P(r)=P_0\sqrt{2}\frac{r}{R}exp[-\frac{r^2}{R^2}+i\tan^{-1}{
(\frac{y}{x})
}],    
\end{equation}
where $R$ is the pumping radius, and $P_0$ is the pumping intensity. Because the external pumping energy is significantly far above the lower polariton branch, the coherent pumping we used was several orders of magnitude smaller than the incoherent pump. Therefore, the number of particles in the coherent reservoir was much smaller compared with the occupation of the incoherent reservoirs in Fig.~\ref{FIG111}(a). Hence, we claim that the macroscopic occupation of the system was due to incoherent pumping. Using the equations above, we investigated the condensate density's dependence on the pumping's size in the pulsed exaction regime An LG pump with duration $t_p=100 ps$ was used.
\begin{equation}
P(r) = P_0\sqrt{2}\frac{r}{R}
exp[-\frac{r^2}{R^2}
+i\tan^{-1}(\frac{y}{x})
]
\theta(t)
\theta(t_p -t)
\end{equation}
The parameters are the same as those used in CW excitation conditions. Fig.~\ref{FIG111}(b) shows the number of condensed particles as a function of time for different pump diameters  (as before, the coherent pump intensity was approximately two orders of magnitude smaller than the incoherent pump intensity). The condensate will form faster if the pumping intensity is fixed and the pump diameter increases from $4.3$ to $6.4 \mu m$.

\begin{figure}[b!]
\centering
\includegraphics[width=1\textwidth]{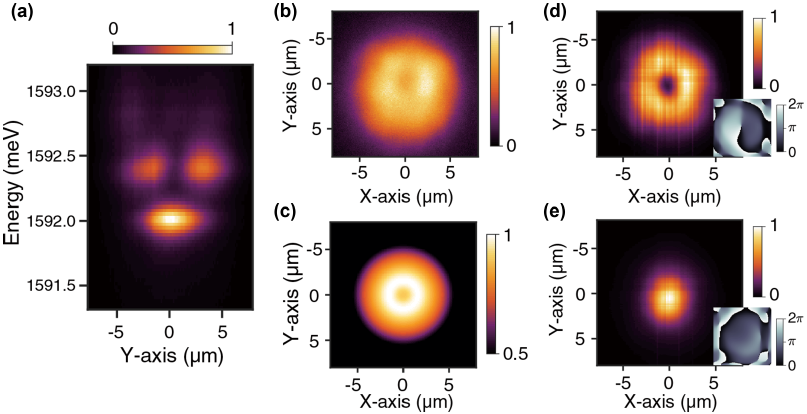}
\caption
[Polariton condensate under continuous wave exciation]
{
The steady state of partially rotating polariton superfluid under CW LG beam excitation. (a) A real-space spectrum of polariton condensates. (b) Energy-integrated PL image of the system. (c) Calculated polariton distribution [simulation of (b)]. Energy-resolved PL images of excited- and ground states are shown in panels (d) and (e), respectively; insets show corresponding phase maps.
}
\label{FIG5}
\end{figure}

The most recent experiment~\cite{PhysRevB.101.245309} describes the decay of vortices in polariton superfluids under net angular-momentum conserving conditions. The result exhibits the simultaneous appearance of the vortices with OAM $l = +1$ and $l = -1$ and the system decay into a purely zero-OAM ground state under the excitation of $l = 0$ ring-shaped pumping. In contrast to their work, our observation shows that a single quantized vortex with a net OAM is spontaneously relaxed in an angular-momentum-nonconserving manner under
an intentional excitation of the $\abs{l} = 1$ LG beam. The reason for the transition in our observation is that a single vortex is generated by nonresonant pump excitation. However, a resonantly created polariton vortex cannot experience such a transition since the resonant pump cannot effectively generate a spatial potential from the excitonic reservoir. Unlike the vortices in an equilibrium superfluid (which do not evaporate over time), the vortices in polariton condensates might decay if the vortex escapes from the system. 

We observed the vanishment of the vortex based on time-resolved measurements of the energy-state transition, which cannot be treated as a direct observation of an escaping vortex. 
We considered this scenario as a distinct possibility, as revealed by our simulations, which show the disappearance of the singularity from the center of the trap. 
Regardless of the actual mechanism through which the vortex escapes from the system, the experiments, conducted by our collaborators, present a knob with which one can control the timing of the switch between the polariton condensates with and without a quantized vortex.

\section{Continuous-wave excitation regime}
For the comparison with pulsed excitation, we also investigated the creation of a polariton vortex under a CW excitation with $l = -1$ OAM beams in the same experimental setup; subsequently, we observed two lowest-energy states Fig~\ref{FIG5}(a). The energy-integrated PL image of the system, depicted in Fig~\ref{FIG5}(b), was consistent with the theoretical simulations in Fig~\ref{FIG5}(c). Fig~\ref{FIG5}(d) and Fig~\ref{FIG5}(e) show the energy-resolved PL images of the excited-
and ground states in the CW excitation regime, respectively. The insets describe the corresponding phase profiles for each state. The main distinction from the pulsed excitation regime is that the CW pump beam
consistently provides the potential barrier, which blocks the excited state from being escaped from the trap. Without diminishing the potential trap during time evolution, the excited state cannot be pushed out of the barrier only by the competition between the two distinct energy states. Therefore, to maintain a steady state, there exists a balance among the decay, pump, and interstate transition processes. We emphasize that this steady state represents an intriguing regime in which a vortex-carrying and ground state coexist.

%%%%%%%%%%%%%%%%%%%%%%%%%%%%%%%%%%%%%%%%%%%%%%%%%%%%%%%%%%%%%%%%%%%%%%%%%%%%%%%%%%%%%%%%%%%%%%%%%5

\section{Summary}
In conclusion, we explicitly demonstrated time-resolved spectroscopy shows a single quantum vortex vanishing in an exciton-polariton superfluid pumped by a pulsed nonresonant LG beam. By using various diameters of the LG pump, we investigated the competition between the vortex-carrying excited state and ground state, which determines the population of the two distinct states of condensates and affects the transition between the states under the pulsed excitation regime. 

The experimental observations are greatly consistent with the numerical calculations using the driven-dissipative Gross-Pitaevskii equation coupled with incoherent and coherent reservoirs. The coherent reservoir gives a rise to the breaking of the rotational symmetry and the revival of the OAM of the nonresonant excitation with the subsequent formation of a quantum vortex with a predefined winding number. 

We proposed experimental and theoretical approaches to manipulate multistate polariton condensates through potential landscape engineering in semiconductor microcavities; these tools have broad implications for investigating light-matter interaction and stability of topological objects in semiconducting devices and developing polariton devices for quantum communications in the near future.
\chapter{Valleytronics of impurity states in two-dimensional Dirac materials}\label{chap:valley}

\section{Introduction}

The critical idea of valleytronics is in using the valley index as an additional active degree of freedom of charge carriers~\cite{PhysRevLett.99.236809,PhysRevLett.108.196802} in gapped graphene~\cite{PhysRevLett.99.236809}, monolayers of transition metal dichalcogenides (TMDs)~\cite{PhysRevLett.108.196802}, among other two-dimensional Dirac materials. 
One of the representatives of TMDs is MoS$_2$: a material with a structure composed of molybdenum atoms sandwiched between pairs of sulfur atoms. 
In contrast to graphene, it is characterized by the inversion symmetry breaking, and it possesses a large band gap with a width in the optical range, absent in monolayer graphene~\cite{Pan2014}. 
It represents a direct band gap material with the minima of the conduction band and maxima of the valence band located at points $K$ and $K'$ in reciprocal space.
Moreover, electrons in MoS$_2$ are subject to strong spin-orbital interaction, which also makes it different from graphene, where the spin-orbital interaction is relatively weak. 
This latter property, due to the electrons occupying d-orbitals in MoS$_2$, results in an extra band splitting~\cite{Silva_Guill_n_2016}.

It has been shown that the interband (between the conduction and valence bands) transitions in Dirac materials are valley-sensitive: at a given circular polarization of the external electromagnetic perturbation, the interband transitions occur predominantly in one valley since the electrons in each valley couple with a specific polarization of light~\cite{Kovalev_2018}. %
Switching to the opposite circular polarization changes the valley where the interband transitions take place~\cite{Zeng_2012}.
These optical selection rules are fulfilled for interband optical transitions, where the electron momentum is a good quantum number. 

However, each material is to some extent disordered: it contains impurities, some of which are unintentional and emerge due to the imperfections of the growth technique, whereas some of the impurities are embedded intentionally to enhance the electronic (or other) properties of the sample.
As a result, additional donor and acceptor energy levels emerge in the vicinity of the conduction and valence bands, respectively.
Then, if the sample is exposed to external light
with the frequency smaller than the band gap of the material, the optical properties become determined mainly by the electron transitions from donor impurities to the conduction band and from the valence band to the acceptor states~\cite{Li1993}. 
In this case, the electron states on impurities are characterized by some quantum numbers instead of the translational momentum due to localization. 
The theoretical description of optical transitions from these states to the bands and the analysis of the corresponding optical selection rules, which consider the valley quantum number, represent an important problem of valley optoelectronics, which has been studied theoretically. 
In particular,  work~\cite{PhysRevLett.121.167402} presents a numerical study of a particular type of disorder: vacancy defects.
%\textcolor{blue}{An addition of atomic vacancies and substitution impurities results in a strain and leads to significant modification of the optical properties of the system~\cite{bahmani2020electronic}.}
%
%
%
\begin{figure}[t!]
\centering
\includegraphics[width=0.8\textwidth]{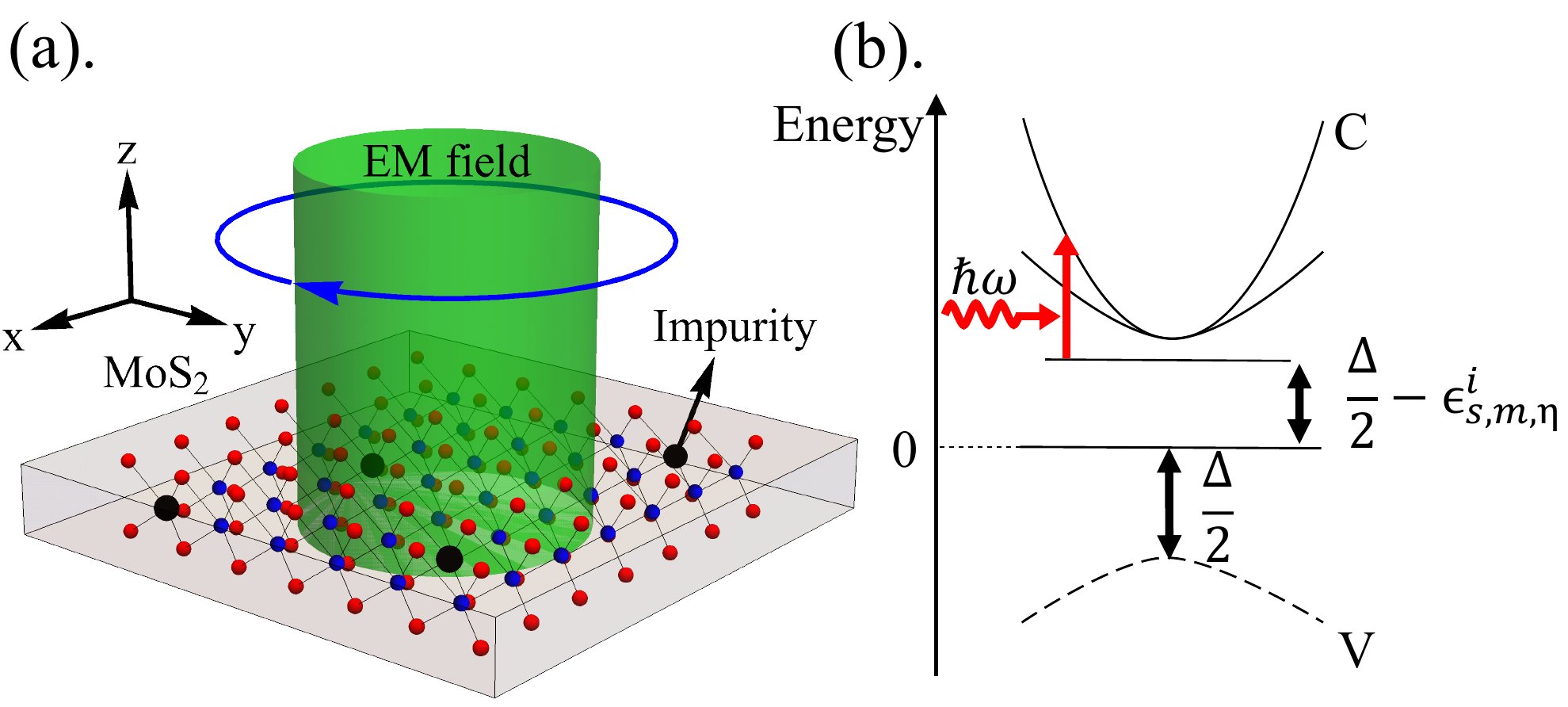}
\caption
[System schematic]
{System schematic. (a) A monolayer of MoS$_2$ (with impurities) exposed to a circularly polarized electromagnetic (EM) field of light (green cylinder). 
(b) The band structure of MoS$_2$. The band gap is $\Delta$, and the system is exposed to an EM field with frequency $\omega$.
}
\label{Fig1}
\end{figure}
%
%
%

%\textcolor{red}{Referee 1-0}
Usually, the energy gap between the impurity states and the conduction band corresponds to the terahertz (THz) frequency range (10-20 meV). 
It can be used to design (pulsed) terahertz radiation detectors. 
A polarized optical signal is transformed into an electric current (to be measured) in such a detector.
The analysis of the optical selection rules here is thus of utmost importance for applications.
We want also to mention another potential application of the theory of impurity-band transitions.
It has been recently proposed that TMD monolayers containing atomic impurities can be used as single-photon emitters~\cite{barthelmi2020atomistic}.
Utilizing artificially-created atomic vacancies, one can achieve the single-photon regime of operation.
This is one of the fast-developing topics of research recently. 

For fundamental purposes and to design the devices, it is useful to study the general properties of defects of any type, and an analytical analysis would be beneficial here.
However, one of the problems to face is that the simple model, which assumes the impurity potential energy to be the Dirac delta-function~\cite{doi:10.1063/1.3556738}, is not applicable in the case of a Dirac Hamiltonian since the electron wave function becomes singular exactly at the center of coordinates~\cite{PhysRevLett.96.126402}.
%This problem has found its  (semi-analytical) solution~\cite{PhysRevLett.96.126402}.
We build a theory of impurity-band transitions in 2D Dirac materials, utilizing and modifying the model of zero-radius impurity potential, which is frequently used for the description of shallow impurities in semiconductors and semiconductor nanostructures~\cite{PhysRevLett.96.126402, pakhomov1996local}. 
We investigate the optical properties of disordered TMDs, examining the light absorption and photoinduced transport effects, accounting for the spin-orbital coupling of electrons.
%We start from the Hamiltonian of an electron bound at the impurity potential with an account of the spin-orbit coupling and subjected to an applied electromagnetic field.
%
We study the behavior of drag electric current density and the absorption coefficient for different vital parameters of the sample and different polarizations of the incident light.
%\textcolor{red}{Referee 2-2(1)}
It should be mentioned that the generation of the electric current in 2D Dirac materials due to the \textit{interband} optical transitions (the photon and phonon drag effects) has been extensively studied~\cite{PhysRevB.81.165441, golub2011valley, PhysRevLett.122.256801, PhysRevB.102.235405, PhysRevB.102.045407}, but the impurity-band transitions have not been addressed.

\section{Hamiltonian and eigenstate}

Light absorption is governed by microscopic transitions of electrons from the
bound impurity states to the conduction band. 
The Hamiltonian of the electron bound at the attractive potential $u(\textbf{r})$ reads
\begin{equation}
\label{EqHam1}
H=\left(\frac{\Delta}{2}\sigma_z + \textbf{v}\cdot\textbf{p}\right)\otimes \id - \frac{\lambda\eta}{2}(\sigma_z-\id)\otimes\hat{s}_z+u(\textbf{r}),
\end{equation}
where $\Delta$ is a band gap, $\textbf{v} = v(\eta\sigma_x,\sigma_y)$ is the velocity operator,
%with $v$ determined by the particular material and having the dimensionality of velocity;
$\textbf{p}$ is the electron momentum, and $\sigma_\alpha$ with $\alpha=x,y,z$ the Pauli matrices of pseudospin. 
The index $\eta=\pm1$ indicates the valley; 
$\lambda$ is intrinsic spin-orbital coupling; $\hat{s}_z$ is the matrix of the electron spin.
The first term in Hamiltonian~(\ref{EqHam1}) %(without the spin-orbit coupling term and the impurity potential terms) 
describes a two-band model of gapped graphene (or a band structure of a TMD material). 
We consider a shallow impurity potential, $u(\textbf{r})$; thus, we assume that the ionization potential of the donor is much smaller than $\Delta$.

To find the eigenfunctions and eigenenergies, we write the Schr\"odinger equation in the momentum representation,
\begin{eqnarray}
\label{EqMatHam}
&&
\begin{pmatrix}
\frac{\Delta}{2}-E & vp_{-} 
\\
vp_{+} & -\frac{\Delta}{2}+s\lambda\eta-E 
\end{pmatrix}
\chi_{s,m}(\textbf{p})
\\
\nonumber
&&~~~~~~~~~~~~~~~~~~~~
+\int\frac{d\textbf{p}'
}{(2\pi\hbar)^2}
u({\textbf{p}-\textbf{p}'})\chi_{s,m}(\textbf{p}')
=0,
\end{eqnarray}
where $s=\pm 1$,  $p_\pm=\eta p_x\pm ip_y=pe^{\pm i\eta\varphi}$ with $\varphi$ the angle of the vector $\textbf{p}$ with respect to $x$-axis, $m$ is the eigenvalue of the z-projection of the electron angular momentum (the quantum number which characterizes electron localized on impurity),
\begin{eqnarray}
&&u(\textbf{p}-\textbf{p}')
=
2\pi\int^\infty_0rdru(r)J_0(|\textbf{p}-\textbf{p}'|r)
\\
\nonumber
&&~~~~~=
2\pi\sum_{k}\int^\infty_0rdru(r)J_k(pr)J_k(p'r)\cos k(\varphi-\varphi '),
\end{eqnarray}
where 
$J_k(x)$ are the $k$-order Bessel functions. 
We search for the spinor eigenfunctions in the form,
\begin{equation}
\label{EqEig1}
\chi_{s,m}(\textbf{p})=\left(
  \begin{array}{c}
    a_{s,m}(p)e^{i m\varphi}\\
    b_{s,m}(p)e^{i(m+\eta)\varphi}
  \end{array}
\right)
\end{equation}
since this form reflects the axial symmetry, 
%of the problem 
and the eigenstates are characterized by the angular momentum projection with quantum number $m$.
Substituting Eq.~\eqref{EqEig1} in Eq.~\eqref{EqMatHam} and performing the integration over $\varphi'$, we find the system of equations for the coefficients $a_{s,m}$ and $b_{s,m}$,
\begin{eqnarray}
\label{Bessel}
&&0=
\begin{pmatrix}
\frac{\Delta}{2}-E & vp
\\
vp & -\frac{\Delta}{2}+s\lambda\eta-E
\end{pmatrix}
\begin{pmatrix}
a_{s,m}(p)\\
b_{s,m}(p)
\end{pmatrix}
\\
\nonumber
&&+
\int\limits^\infty_0rdru(r)\int\limits^\infty_0
\frac{p'dp'}{\hbar^2}
\begin{pmatrix}
J_m(pr)J_m(p'r)a_{s,m}(p')\\
J_{(m+\eta)}(pr)J_{(m+\eta)}(p'r)b_{s,m}(p')
\end{pmatrix}.
\end{eqnarray}
%
%This equation is still exact. 
To draw principal conclusions, we can now simplify these equations.
Only the low-lying impurity states are occupied for a shallow impurity ($\epsilon^i_{s,m,\eta}\ll\Delta$) and low enough temperatures. Then, we can consider the transitions from impurity states corresponding to $m=0$ and $m=\pm1$ levels only.
Assuming that the potential of each impurity $u(r)$ is sharply peaked in the vicinity of its center $r=0$ and rapidly decreases with $r$, we can take the Bessel functions under the integral $r=0$. For the $m=0$ state, $J_0(0)=1$ and $J_{\eta=\pm1}(0)=0,$ and we find the simplified form of Eq.~\eqref{Bessel},
\begin{equation}
\label{simple}
\begin{split}
\begin{pmatrix}
\epsilon^i_{s,0,\eta} & vp \\
vp & -\Delta+s\lambda\eta
\end{pmatrix}
\begin{pmatrix}
a_{s,0} \\ b_{s,0}
\end{pmatrix}
+
\begin{pmatrix}
Au_0 \\ 0
\end{pmatrix}
=0,
\end{split}    
\end{equation}
where 
\begin{equation}
\label{Parms}
A=\int^\infty_0\frac{p'dp'}{\hbar^2}a_{s,0}(p');~~~ u_0 = \int^\infty_0u(r)rdr.
\end{equation}
The solution of ~Eq .\eqref{simple} reads 
\begin{equation}\label{WF}
\begin{pmatrix}
a_{s,0} \\ b_{s,0}
\end{pmatrix}
=-
\hbar v
\sqrt{\frac{2\pi\epsilon^i_{s,0,\eta}}{\Delta}}
\begin{pmatrix}
\frac{\Delta-s\lambda\eta}{(vp)^2+\epsilon^i_{s,0,\eta}(\Delta-s\lambda\eta)}
\\
\frac{vp}{(vp)^2+\epsilon^i_{s,0,\eta}(\Delta-s\lambda\eta)}
\end{pmatrix}.
\end{equation}
%
%!!!
%!!!
%!!!
%Combining Eq.~\eqref{Parms} and Eq.~\eqref{WF} we find the dispersion equation for energy levels $\epsilon^i_{s,0,\eta}$.
%------------

%\subsubsection{$m = 1$ state}
For the $m = 1, \eta = -1$ state, 
$J_0(0)=1$, $J_1(0)=0$, and $J_2(0)=0$,  Eq.~\eqref{Bessel} can be simplified,
%We will also assume a shallow donor center. Thus we take $0<\epsilon_i\ll\frac{\Delta}{2}$ and find
%
\begin{equation}\label{M1eq}
\begin{split}
\begin{pmatrix}
\epsilon^i_{s,1,-1} & vp \\
vp & -\Delta-s\lambda
\end{pmatrix}
\begin{pmatrix}
a_{s,1} \\ b_{s,1}
\end{pmatrix}
+
\begin{pmatrix}
0 \\ Bu_0
\end{pmatrix}
=0,
\end{split}    
\end{equation}
where now
\begin{equation}
\label{defB}
B=\int^\infty_0\frac{p'dp'}{\hbar^2}b_{s,1}(p').
\end{equation}
The solution of ~Eq. \eqref{M1eq} reads
\begin{equation}
\begin{pmatrix}
a_{s,1} \\ b_{s,1}
\end{pmatrix}
=-
\hbar v
\sqrt
{
\frac
{2\pi(\Delta^2-\lambda^2)}
{\Delta\epsilon^i_{s,1,-1}}
}
\begin{pmatrix}
\frac{vp}{(vp)^2+\epsilon^i_{s,1,-1}(\Delta+s\lambda)}
\\
\frac{-\epsilon^i_{s,1,-1}}{(vp)^2+\epsilon^i_{s,1,-1}(\Delta+s\lambda)}
\end{pmatrix}.
\end{equation}
The energy $\epsilon^i_{s,1,-1}$ of this state can also be found using the definition Eq.~\eqref{defB}. We see that within the framework of the shallow-impurity model, the state $m=1$ forms `under' the $\eta=-1$ valley (and vice versa).
In other words, the following rule holds $m+\eta=0$ for $m=\pm1$ states.
%This wave function describes the impurity states created by the impurity potential. 
%The impurity state is degenerated by the valley index $\eta$. 
%To find the splitting of impurity states by the valley index, one needs to keep nonzero terms when we approximate the Bessel functions $J_{\eta=±1}(x)$ at small values of its arguments. Further, we will assume that the valley index degenerates the impurity state and ignore this small splitting.

The electron states in the wave function describes the conduction band,
%
% \begin{equation}
% \psi_{s,\eta}(\textbf{p})=
% \begin{pmatrix}
% \cos\left(\frac{\theta_{s,\eta}}{2}\right), 
% &
% %\\
% \sin\left(\frac{\theta_{s,\eta}}{2}\right)e^{i\eta\varphi_{\textbf{p}}}
% \end{pmatrix}^\mathrm{T},
% \end{equation}
\begin{equation}
\psi_{s,\eta}(\textbf{p})=\left(
  \begin{array}{c}
   \cos\left(\frac{\theta_{s,\eta}}{2}\right) \\
   \sin\left(\frac{\theta_{s,\eta}}{2}\right)e^{i\eta\varphi_{\textbf{p}}}
  \end{array}
\right),
\end{equation}
where we use the notations $\cos\theta_{s,\eta}=(\Delta-s\lambda\eta)/2E_{s,\eta}(\textbf{p})$ and $\sin\theta_{s,\eta}=\eta vp/E_{s,\eta}(\textbf{p})$ 
with the conduction band electron energy $E_c(\textbf{p}) = s\eta\lambda/2 + E_{s,\eta}(\textbf{p})$, $E_{s,\eta}(\textbf{p})=\sqrt{(vp)^2+\left[(\Delta-s\lambda\eta)/2\right]^2}$.
Since the transitions from the impurity state with a given valley number $\eta$ to the conduction band of the other valley $\eta'\neq\eta$ are strongly suppressed due to the significant distance between the valleys in the reciprocal space~\cite{Gorycaeaau4899}, the main contribution to the light absorption comes from the impurity-band transitions with the same valley number $\eta' = \eta$. 
The circularly polarized EM field case is the most interesting from the point of view of applications. 
The Hamiltonian describing the interaction of electrons with the external EM perturbation reads $\hat{V}(\textbf{}{r},t)=-e\textbf{v}\cdot\mathbf{A}(\textbf{r},t)$,
%
%\begin{equation}
%\hat{V}=-e\textbf{v}\cdot\mathbf{A}(t),    
%\end{equation}
%
where $\mathbf{A}(\mathbf{r},t) = \mathbf{A}_0 exp(i\mathbf{kr}-i\omega t) + \mathbf{A}^{*}_0 exp(-i\mathbf{kr}+i\omega t)$ is the vector potential of EM field.
%\textcolor{red}{Referee 1-1}
Here, $\textbf{A}_0 = A_0\hat{x}+A_0i\sigma\hat{y}$ with $\sigma$ the light polarization, $\hat{x}$ and $\hat{y}$ the unity vectors in the corresponding directions in direct space, $\mathbf{k}$ and $\omega$ being the photon wave vector and frequency, respectively.

\section{Electric current density}
\begin{figure}[t!]
\centering
\includegraphics[width=0.7\textwidth]{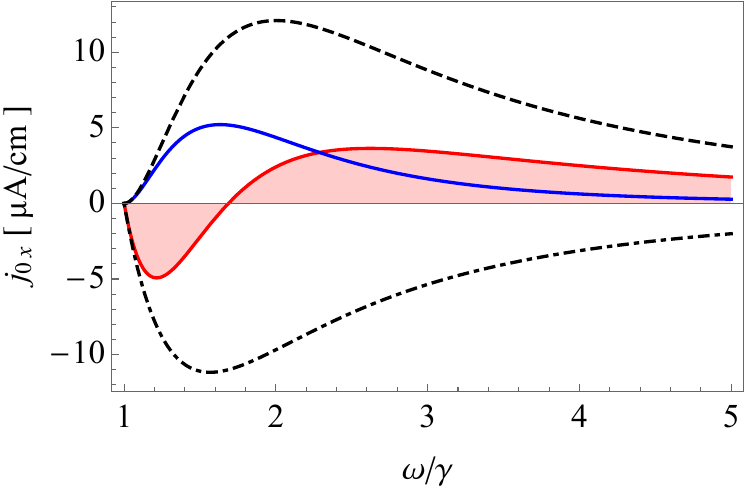}
\caption
[Current density $j_{0x}$]
{Spectrum of electric current density due to the transitions from $m=0$, $\eta=-1$ 
impurity states to spin-up conduction band states for the polarization of light $\sigma = -1$ (red) and $\sigma = 1$ (blue); $\gamma=\epsilon^{i}_{1,0,-1}/\hbar$, 
where $\epsilon^{i}_{1,0,-1} = 10$~meV is the energy of impurity counted from the bottom of the conduction band.
Black curves show the positive and negative contributions to the current in $\sigma = -1$ case.
We used the density of impurities
$n_i \approx 5\times10^{12}$~cm$^{-2}$; 
%for the impurity estimating it by the formula $n_i\sim 1\%$ of $\rho/m_a$, where $\rho=5060$~kg$/$m$^3$ is the density of MoS$_2$, $m_a= 3\times10^5$~$m_0$ the mass of a MoS$_2$ molecule;
electron relaxation time $\tau=2\times10^{-13}$~s, velocity $v=at/\hbar$, the lattice constant of MoS$_{2}$ $a=3.193$~\AA,
effective hopping integral $t=1.10$~eV~\cite{Hatami_2014}, amplitude of light
$A_0=3.8\times10^{12}$ eV$\cdot$ s/C$\cdot m$, the band gap $\Delta=1.16$~eV, and the spin-orbit coupling strength $\lambda=75$~meV.
}
\label{Fig2}
\end{figure}

The general expression for the (partial) component of photon-drag electric current density, corresponding to electron transitions from the impurity state with a quantum number $m$ to the conduction band, reads $(\alpha = x, y)$
\begin{equation}
\label{EqCurrentMain}
j_{m\alpha}=\frac{2\pi en_i\tau}{\hbar}\int
\frac{v_\alpha(\textbf{p})d\textbf{p}}{(2\pi\hbar)^2}
|M_m(\textbf{p},\textbf{k})|^2
\delta(E_{c}(\textbf{p})-E_i-\hbar \omega),
\end{equation}
where $n_i$ is the impurity concentration, $e$ is the elementary charge, $\tau$ is the electron relaxation time in the conduction band, $M_m(\mathbf{p},\mathbf{k})=\langle\psi_{s,\eta}(\mathbf{p})|\hat{V}|\chi_{s,m}(\mathbf{p}-\mathbf{k})\rangle$ is the matrix element of impurity-band transitions, and the electron velocity components read $v_x(\textbf{p})=\eta\sin\theta_\textbf{p}\cos\varphi_\textbf{p}$ and $v_y(\textbf{p})=-i\eta\sin\theta_\textbf{p}\sin\varphi_\textbf{p}$; 
$E_i = \Delta/2 - \epsilon^i_{s,m,\eta}$ is the impurity energy level. The detailed calculations for electric current density are discussed in appendix~\ref{current_density_appendix}
%
%\begin{equation}
%\begin{split}
%v_x(\textbf{p})=\eta\sin\theta_\textbf{p}\cos\varphi_\textbf{p};\,\,\,\,
%v_y(\textbf{p})=-i\eta\sin\theta_\textbf{p%}\sin\varphi_\textbf{p}.
%\end{split}    
%\end{equation}
%
%\subsubsection{m = 0}
%
%
%0.385
%
%
%
\begin{figure}[t!]
\centering
\includegraphics[width=0.7\textwidth]{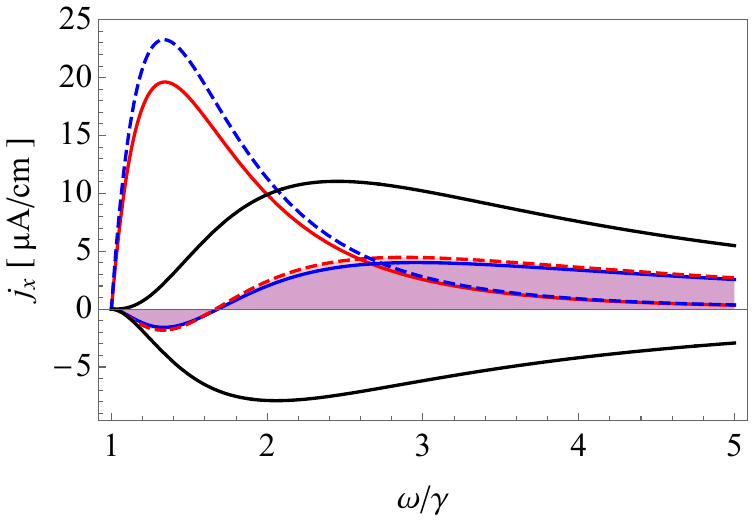}
\caption
[Current density $j_{1x}$]
{Spectrum of electric current density due to the transitions from $m=1$, $\eta=-1$ (solid) and $m=-1$, $\eta=1$ (dashed) impurity states to spin-up conduction band states for the polarizations of light $\sigma = -1$ (red) and $\sigma = 1$ (blue); $\gamma=\epsilon^{i}_{1,1,-1}/\hbar$.
The other parameters are taken the same as in Fig.~\ref{Fig2}. }
\label{Fig3}
\end{figure}

Considering  the $m = 0$ impurity state, we find the corresponding matrix element,
\begin{eqnarray}
\label{EqM0}
|M_0(\textbf{p,k})|^2 &=&
(evA_0)^2
(\hbar v)^2
\frac
{2\pi\epsilon^i_{s,0,\eta}}
{\Delta}
\\
\nonumber
&&
\times
\left\{
\frac{(\eta+\sigma)^2v^2(\textbf{p}-\textbf{k})^2\cos^2\left(\frac{\theta_{s,\eta}}{2}\right)
}{\Big[v^2(\textbf{p}-\textbf{k})^2+\epsilon^i_{s,0,\eta}(\Delta-s\lambda\eta)\Big]^2}
\right.
\nonumber
% \\
% \nonumber
% &&~~~~~~~
\left.
+
\frac{
(\eta-\sigma)^2(\Delta-s\lambda\eta)^2\sin^2\left(\frac{\theta_{s,\eta}}{2}\right)
}{\Big[v^2(\textbf{p}-\textbf{k})^2+\epsilon^i_{s,0,\eta}(\Delta-s\lambda\eta)\Big]^2}
\right\}.
\end{eqnarray}
Without the loss of generality, let us choose $\mathbf{k}$ to be directed along the x-axis ($\mathbf{k}=k\hat{x}$). 
Then, in Eq.~\eqref{EqCurrentMain} only the term containing $\cos\varphi_\textbf{p}$ survives and $j_y = 0$, which reflects the fact that the photon-drag current should be directed along the photon wave vector. 
%The electric current density can be split into two components, $j = j_1 + j_2$,
Substituting Eq.~\eqref{EqM0} in Eq.~\eqref{EqCurrentMain}, we find
\begin{equation}
\begin{split}
&
j_{0x}
=
\beta_{0}'\Theta[\delta\hbar \omega_{s,0,\eta}]
\frac{k\pi}{v}\frac{(\Delta-s\lambda\eta)+\delta\hbar \omega_{s,0,\eta}}{\Big[(\delta\hbar \omega_{s,0,\eta})^2 + (\Delta - s\lambda\eta)\hbar \omega\Big]^2}
\\
&
\times
\frac{\delta\hbar \omega_{s,0,\eta}}{(\Delta - s\lambda\eta) + \delta\hbar \omega_{s,0,\eta}}
 \Bigg\{
\frac{4(\Delta-s\lambda\eta)^2}{(\delta\hbar \omega_{s,0,\eta})^2 + (\Delta - s\lambda\eta)\hbar \omega}
\\
&
\times
\delta\hbar \omega_{s,0,\eta}(\eta-\sigma)^2
+
\Big[(\Delta-s\lambda\eta)+\delta\hbar \omega_{s,0,\eta}\Big](\eta+\sigma)^2
\\
\label{EqCurrent0}
&
\times
\left[
\frac{4\Big((\Delta-s\lambda\eta)+\delta\hbar \omega_{s,0,\eta}\Big)\delta\hbar \omega_{s,0,\eta}}{(\delta\hbar \omega_{s,0,\eta})^2 +(\Delta-s\lambda\eta)\hbar \omega}
-2\right]
\Bigg\},
\end{split}
\end{equation}
where 
$\beta_{0}'=en_i\tau v^2\epsilon_i(evA_0)^2/\hbar\Delta$ and $\delta\hbar\omega_{s,m,\eta}=\hbar\omega-\epsilon^i_{s,m,\eta}$.
Figure~\ref{Fig2} shows the spectrum of electric current density for different polarizations of light and $m=0$, $\eta=-1$.
Interesting to note that in the case of the polarization of light $\sigma = -1$, the electric current flows in the opposite direction in some regions of frequencies, and then it changes its direction. 
%\textcolor{red}{Referee 2-2(2)}
A similar behavior (of the inversion of the direction of the electric current density) was demonstrated in work~\cite{PhysRevB.81.165441}.
Mathematically, it happens due to an interplay of different terms in Eq.~\eqref{EqCurrent0}, shown as dashed curves.
Such behavior is not observed for $\sigma=1$. For the case $m = 1$ (and, correspondingly, $\eta=-1$), we find
\begin{eqnarray}
\label{EqM1}
|M_{1}(\textbf{p},\textbf{k})|^2&=&
(evA_0)^2
(\hbar v)^2
\frac
{2\pi(\Delta^2-\lambda^2)}
{\Delta\epsilon^i_{s,1,-1}}
\\
\nonumber
&&\times
\left\{
\frac{(\sigma-1)^2(\epsilon^i_{s,1,-1})^2\cos^2\left(\frac{\theta_{s,-1}}{2}\right)}{\Big[v^2(\textbf{p}-\textbf{k})^2+\epsilon^i_{s,1,-1}(\Delta+s\lambda)\Big]^2}\right.
\nonumber
+\left.
\frac{(\sigma+1)^2v^2(\textbf{p}-\textbf{k})^2\sin^2\left(\frac{\theta_{s,-1}}{2}\right)}{\Big[v^2(\textbf{p}-\textbf{k})^2+\epsilon^i_{s,1,-1}(\Delta+s\lambda)\Big]^2}
\right\}.
\end{eqnarray}
%
%The current density aligned in the x-direction can be split into two contributions $j_x=j_1+j_2$, %
%\begin{equation}
%\begin{split}
%&j_1 = \alpha'\int\frac{vp}{E_s}\cos\varphi d\textbf{p}\frac{(\eta+\sigma)^2\epsilon_i^2}{\Big[\Big(v(\textbf{p}-\textbf{k})\Big)^2+\epsilon_i(\Delta-\lambda\eta)\Big]^2}\cos^2(\frac{\theta_s}{2})\\&\times\delta(E_s-E_i-\hbar\omega),\\
%&j_2 = \alpha'\int\frac{vp}{E_s}\cos\varphi d\textbf{p}\frac{(\eta-\sigma)^2v^2(\textbf{p}-\textbf{k})^2}{\Big[\Big(v(\textbf{p}-\textbf{k})\Big)^2+\epsilon_i(\Delta-\lambda\eta)\Big]^2}\sin^2(\frac{\theta_s}{2})\\&\times\delta(E_s-E_i-\hbar \omega),\end{split}    
%\end{equation}
%
%Calculating the integrals, we find
Again, only the x-component of the current is finite,
\begin{equation}
\label{EqCurrent1}
\begin{split}
j_{1x}
=
\beta_1'\Theta[\delta\hbar\omega_{s,1,-1}]\frac{k\pi}{v}\frac{(\Delta+s\lambda)+\delta\hbar\omega_{s,1,-1}}{\Big((\delta\hbar\omega_{s,1,-1})^2 + (\Delta + s\lambda)\hbar \omega\Big)^2}
\\
\times
\frac{\delta\hbar\omega_{s,1,-1}}{(\Delta+s\lambda) + 2\delta\hbar\omega_{s,1,-1}}
 \Bigg\{
\frac{(\Delta+s\lambda)+\delta\hbar\omega_{s,1,-1}}{(\delta\hbar\omega_{s,1,-1})^2 + (\Delta + s\lambda)\hbar \omega}
\\
\times
4(\epsilon{^i_{s,1,-1}})^2(\sigma-1)^2+\delta\hbar\omega_{s,1,-1}(\sigma+1)^2
\\
\times
\left(
\frac{4\Big((\Delta+s\lambda)+\delta\hbar\omega_{s,1,-1}\Big)\delta\hbar\omega_{s,1,-1}}{(\delta\hbar\omega_{s,1,-1})^2 +(\Delta+s\lambda)\hbar \omega}
-2\right)
 \Bigg\},
\end{split}
\end{equation}
where $\beta_{1}'=\beta_{0}'[\Delta^2-\lambda^2]/\epsilon_i^2$. 

Figure~\ref{Fig3} shows the spectrum of electric current density for $m = -1$ and $m = 1$. 
We choose $\eta = -1$ for $m = 1$ and $\eta = 1$ for $m = -1$ .
%since for $\eta=1$ the current is zero here (compare with Fig.~\ref{Fig2}).
Also, the $\sigma=1$ polarization gives the region of positive and negative electric currents for $m = 1$ (blue solid curve). For $m = -1$, the $\sigma = -1$ polarized light gives such electric current (red dashed curve).
For a given $\sigma$, we have optical transitions in both the K (for $m=0$ or 1) and K$'$ (for $m=1$ or 0) valleys. 
They can sum up or partially compensate each other.

\section{Symmetry analysis of the electric current density}

Let us now analyze the formulas for the electric current density [Eqs.~\eqref{EqCurrent0} and~\eqref{EqCurrent1}] from the symmetry point of view and compare them with the case of a graphene monolayer~\cite{glazov2014high}.
Single-layer graphene (without a substrate) possesses the $D_{6h}$ point group, while single-layer MoS$_2$ has $D_{3h}$ point group. 
However, for both the groups, the fourth-rank (generalized conductivity) tensor $\Phi_{\alpha\beta\gamma\mu}$ is the same~\cite{boyd2020nonlinear}.

The general expression for the electric current density reads~\cite{glazov2014high}
\begin{equation}
\label{EqCur1x}
j_{x} = T_1 k_x \frac{|E_x|^2 + |E_y|^2}{2} +
T_2 k_x \frac{|E_x|^2 - |E_y|^2}{2},
\end{equation}
where $E_x$ and $E_y$ are the components of the electric field; 
$T_1$ and $T_2$ are constants describing linear photon drag effect. 
Since the electric field is circularly polarized in our case, $\mathbf{E}=E_0(1,i)$,
only the first term in Eq.~\eqref{EqCur1x} remains, and the other one vanishes since $|E_x|=|E_y|$. 
We see that Eqs.~\eqref{EqCurrent0} and~\eqref{EqCurrent1} obey the symmetry properties of the system.

\section{Light absorption coefficient}
Furthermore, let us study the light absorption coefficient for the $m-$th impurity state.
It is defined as the ratio of the energy flux of absorbed photons and the average energy flux of incident photons~\cite{fang2013quantum}, $\alpha_m(\hbar \omega) = \hbar \omega W_m/P$,
where $P$ are the average of the Poynting flux for the light intensity~\cite{chuang2012physics}, $P=n_rc\epsilon_0\omega^2A_0^2/2$, 
where $n_r$ is the refractive index of MoS$_2$ and $\epsilon_0$ is the vacuum permittivity.
The probability of light absorption in a given valley $\eta$ and from a particular impurity state $m$ is given by the Fermi golden rule,
\begin{equation}
\label{PROBA}
W_m(\omega)=\frac{2\pi n_i}{\hbar}\int\frac{d\textbf{p}}{(2\pi\hbar)^2}|M_m(\textbf{p},0)|^2\delta(E_{c}(\textbf{p})-E_i-\hbar \omega).
\end{equation}
For the transition from $m=0$ impurity state, we find
\begin{equation}
\begin{split}
&\alpha_0 = 
\frac{2\pi n_ie^2v^2\epsilon^i_{s,0,\eta}}{n_rc\omega\Delta\epsilon_0}
\Theta[\delta\hbar \omega_{s,0,\eta}]
\frac{\delta\hbar \omega_{s,0,\eta}}{(\Delta - s\lambda\eta)+2\delta\hbar \omega_{s,0,\eta}}
~~~
\\
&\times
\frac{\sqrt{4\Big(\Delta-s\lambda\eta+\delta\hbar \omega_{s,0,\eta}\Big)\delta\hbar \omega_{s,0,\eta}+(\Delta-s\lambda\eta)^2}}{\Big((\delta\hbar \omega_{s,0,\eta})
 ^2+(\Delta-s\lambda\eta-\epsilon^i_{s,0,\eta})\hbar \omega\Big)^2}
 %~~~~~~~~~~~~~~~~~~~~
\\
&\times
\left\{
(\eta+\sigma)^2\Big(\Delta-s\lambda\eta+\delta\hbar \omega_{s,0,\eta}\Big)^2\right.
\nonumber
\left.
 +(\eta-\sigma)^2(\Delta - s\lambda\eta)^2
 \right\},
\end{split}
\end{equation}
and for $m=1,\eta=-1$ state,
\begin{equation}
\begin{split}
\alpha_1 & = 
\frac{2\pi n_ie^2v^2}{n_rcw\Delta\epsilon_0}
\Theta[\delta\hbar \omega_{s,1,-1}]
\frac{\Delta^2-\lambda^2}{\epsilon^i_{s,1,-1}}
%
%\Theta[\hbar \omega - \epsilon^i_{s,1,-1}]
%
\frac{(\Delta+s\lambda)+\delta\hbar \omega_{s,1,-1}}{(\Delta + s\lambda)+2\delta\hbar \omega_{s,1,-1}}
\\\times&
\frac{\sqrt{4\Big(\Delta+s\lambda+\delta\hbar \omega_{s,1,-1}\Big)\delta\hbar \omega_{s,1,-1}+(\Delta+s\lambda)^2}}{\Big((\delta\hbar \omega_{s,1,-1})
 ^2+(\Delta+s\lambda-\epsilon^i_{s,1,-1})\hbar \omega\Big)^2}
\\\times&
\left\{
(\sigma-1)^2(\epsilon^i_{s,1,-1})^2
 +(\sigma+1)^2(\delta\hbar \omega_{s,1,-1})^2
 \right\}.
 %\Theta[\delta\hbar \omega_{s,1,-1}]
\end{split}
\end{equation}
\begin{figure}[t!]
\centering
\includegraphics[width=0.7\textwidth]{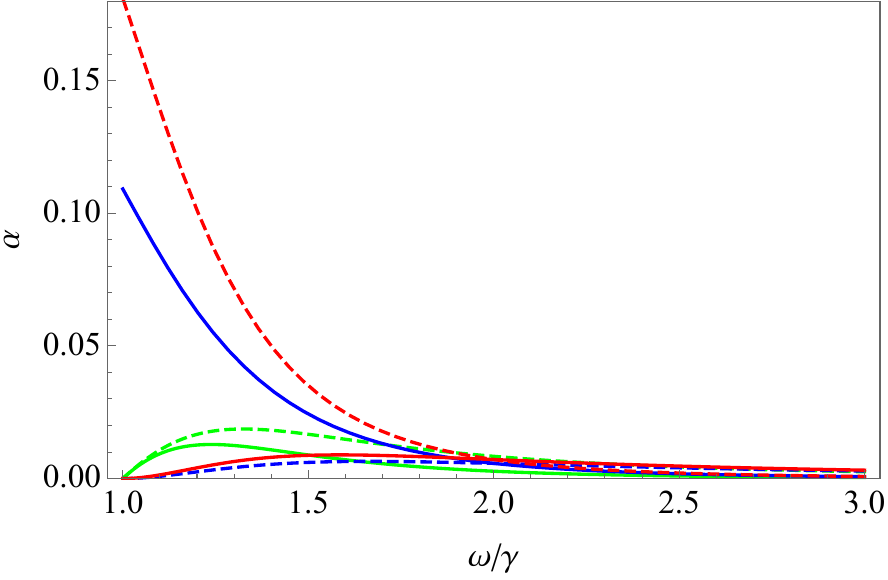}
\caption
[Absorbance spectrum]
{Spectrum of
absorbance for $m = -1$ (red), $m = 0$ (green), and $m = 1$ (blue); $\sigma = 1$ (dashed) and $\sigma = -1$ (solid).
}
\label{Fig4}
\end{figure}
%
%---------------
%---------------
%---------------
%

Figure~\ref{Fig4} shows the spectra of absorbance. 
%for the cases  $m = 0$ and $m=1$. 
%%%!!! Can be removed below!
%It demonstrates valley-dependant optical selection rules. 
For the transitions from the state $m = 1$, the magnitude of absorbance is higher for the $\sigma = -1$ light, but the valley dependence disappears by increasing the photon energy.
For the transitions from the state $m = 0$, both the polarized lights give a comparable contribution.
%
%\section{conclusions}

%\textcolor{red}{Referee 1-3}
It is enlightening to compare the matrix element corresponding to impurity-band transitions with the matrix element for the interband transitions, $|M_{cv}(\mathbf{}{p})|^2$~\cite{Kovalev_2018, PhysRevB.103.035434}. 
The valley selectivity for interband transitions is, to a large extent, satisfied only at small values of momentum $p$, giving $|M_{cv}(0)|^2\propto(\eta+\sigma)^2$. 
In our case, the transitions from $m=0$ impurity states are strongly suppressed due to $|M_0(0,0)|^2\rightarrow0$, whereas for $m=\pm1$ we find $|M_{m=\pm1}(0,0)|^2\propto\epsilon_i^2(\sigma+\eta)^2$ under the condition $m+\eta=0$. 
It means that the valley selectivity takes place for orbital impurity states $m=\pm1$ (and thus, we have $\exp(im\varphi)\neq1$), reflecting the chirality of the band electron wavefunction. These general conclusions are supported by numerical analysis. 
For instance, the absorption coefficient in Fig.~\ref{Fig4} is large in the vicinity of the threshold for $m=1,\eta=-1$ state at $\sigma=-1$ polarization.

\section{Summary}
We 
have studied the selection rules for the light-induced transitions from impurity states to the conduction band in two-dimensional gapped Dirac materials.
We calculated and investigated
the absorption coefficient 
and the photon-drag-induced electric current.
For clarity, we used the shallow impurity potential model.  
Nevertheless, this model correctly reflects the selection rules of any impurity possessing azimuthal symmetry. 
Thus, our conclusions on the optical selection rules are sufficiently general. 
\chapter{Summary}\label{chap:summary}
Chapter 1 is devoted to a general introduction to exciton-polaritons. We discussed the formation of excitons in bulk semiconductors, quantum well, and transition metal dichalcogenide materials. We derived the dispersion of exciton-polaritons and the formation of polariton condensates with a brief description of the history in experiments and numerical modeling to simulate the appearance and evolution of polariton condensates. 

In chapter 2, we studied the partial quantum revivals of compact localized states loaded in a distorted kagome lattice. In the tight-binding model, the reappearance of the initial compact-localized states took place in the neighboring sites of the lattice, but the revival of the original compact-localized states is not observed. In contrast, in the continuous model, apparent revivals of the polariton compact-localized states were observed by squeezing the lattice in either a vertical or horizontal direction which deforms the nearly-flat band. 

In chapter 3, we demonstrated a quantum polariton vortex vanishment excited by a ring-shaped Laguerre-Gaussian pump. The competition between ground and vortex states was observed in pulsed excitation by changing the diameter of the Laguerre-Gaussian pump. It affected the transition between the two states and determines the multistate condensate population. For the simulation, we utilized the driven-dissipative Gross-Pitaevskii equation coupled with incoherent and coherent reservoirs. An external pumping source directly provided particles in the inactive reservoir. Particles in the active reservoir were refilled from the inactive reservoir and then the polariton condensate was replenished by the active reservoir We showed that the numerical simulation was consistent with the experimental results.

In chapter 4,  we studied the valley selection rules for the light-induced transitions from impurity states to the conduction band in two-dimensional gapped Dirac materials. We calculated the absorption coefficient and the photon-drag-induced electric current for the various transitions with different light polarizations. The electric current induced by the transition from impurities showed the flow of opposite direction in some frequency regions with specific polarization of light. For the transitions from the impurity state
$m = 1$, the absorption coefficient had a higher magnitude for the $\sigma = -1$ light, but the valley dependence vanished by increasing the frequency of the light. However, with the impurity state $m = 0$, both the polarized lights showed comparable behaviors.
\chapter{Appendix}\label{chap:Appendix}

\section{Dynamics of polariton compact localized states in kagome x-squeezed strip}\label{cls_appendix}

\begin{figure}[b!]
\centering
\includegraphics[width=0.7\textwidth]{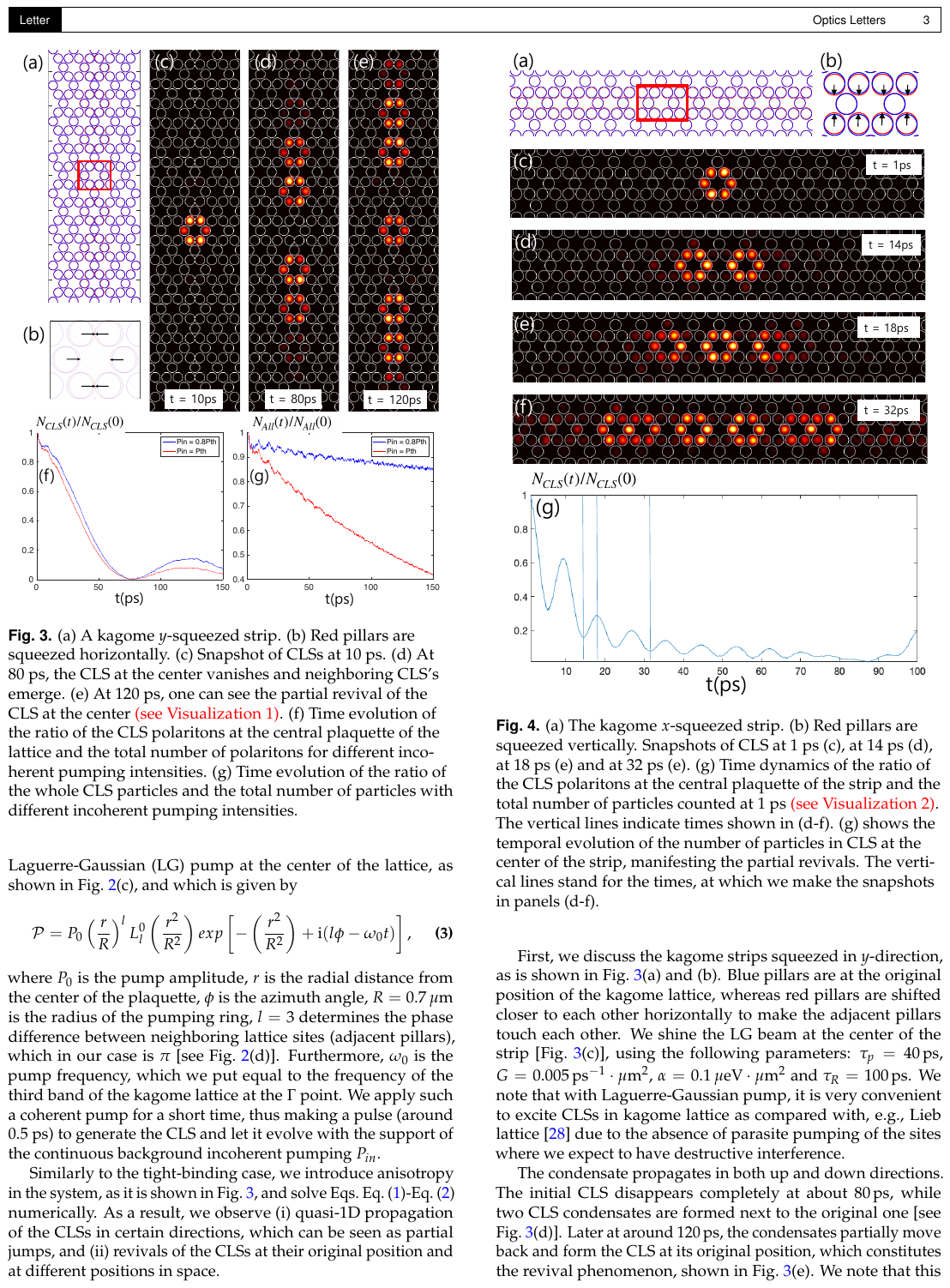}
\caption
[Kagome $x$-squeezed strip]
{
(a) A kagome $x$-squeezed strip. (b) Red pillars are squeezed horizontally. (c) Snapshot of CLSs at 10~ps. (d) At 80~ps, the CLS at the center vanishes, and neighboring CLS's emerge. (e) At 120~ps, one can see the partial revival of the CLS at the center. (f) Time evolution of the ratio of the CLS polaritons at the central plaquette of the lattice and the total number of polaritons for different incoherent pumping intensities. (g) Time evolution of the ratio of the whole CLS particles and the total number of particles with different incoherent pumping intensities.
}
\label{fig3}
\end{figure}

We discuss the kagome strips squeezed in $x$-direction,
% Revival of the CLS on Kagome Y strips are presented in FIG3.(b)~(d). FIG. 3.(a) shows the anisotropic kagome Y strips. 
as is shown in Fig.~\ref{fig3}(a) and (b). Blue pillars are at the original position of the kagome lattice, whereas red pillars are shifted closer to each other horizontally to make the adjacent pillars touch each other.
We illuminate the LG beam at the center of the strip [Fig.~\ref{fig3}(c)], using the following parameters:
$\tau_{p}=40\,\mathrm{ps}$, $G=0.005\,\mathrm{ps}^{-1}\cdot\mu\mathrm{m}^2$,  
$\alpha=0.1\,\mu\mathrm{eV}\cdot\mu\mathrm{m}^{2}$ and $\tau_{R}=100\,\mathrm{ps}$. 
We note that with the Laguerre-Gaussian pump, it is very convenient to excite CLSs in kagome lattice compared with, e.g., Lieb lattice~\cite{PhysRevB.98.161204} due to the absence of parasite pumping of the sites where we expect to have destructive interference.

The condensate propagates in both up and down directions. The initial CLS disappears completely at about $80\,\mathrm{ps}$, while two CLS condensates are formed next to the original one [see Fig.~\ref{fig3}(d)].
Later at around $120\,\mathrm{ps}$, the condensates partially move back and form the CLS at its original position, which constitutes the revival phenomenon, shown in Fig.~\ref{fig3}(e).
We note that this is a partial revival since the whole state of the quantum system does not fully coincide with its original state, and it is characterized by the presence of a few CLS condensates further away from the creation center.

Figure~\ref{fig3}(f) shows the dynamics of the number of polaritons in the original CLS [at the center of the strip, as in~\ref{fig3}(c)] as a ratio of the total number of particles in the whole strip for different incoherent pumping intensities, $P_{in}$. 
By increasing $P_{in}$, the number of particles in the strip also increases.
This figure also provides additional confirmation of the revival effect.

Figure~\ref{fig3}(g) shows the ratio of the total number of polaritons in the strip (consisting of nine plaquettes) and the total number of particles in the system.
If $P_{in}$ is less than the pumping threshold $P_{th}=(G\tau_{R}\tau_{P})^{-1}$, the total number of particles decays.
When $P_{in}=P_{th}$, the losses due to the finite lifetime become saturated due to the compensation by the incoherent pumping reservoir.
Nevertheless, this background incoherent pump does not hinder the phenomenon we are interested in: the revival of the CLS persists, as it is shown in Fig.~\ref{fig3}(c-e) with the condition $P_{in}=P_{th}$.

\newpage
\section{Detailed calculation of $j_{0x}$ and $j_{1x}$} \label{current_density_appendix}
\subsection{Derivation of $j_{0x}$} \label{j_0x_appendix}
To show the detailed derivation of $j_{0x}$, we start from the matrix element $M_0(\textbf{p},\textbf{k})$ in Eq.~\ref{EqM0}.
\begin{align}
%M_p
%%%%%%%%%%%%%%% 1 %%%%%%%%%%%%%%%
M_0(\textbf{p},\textbf{k}) &= \bra{\psi_{s,\eta}(\textbf{p})}\hat{V}\ket{\chi_{s,0}(\textbf{p}-\textbf{k})}     
\nonumber
\\\nonumber
% %%%%%%%%%%%%%%% 2 %%%%%%%%%%%%%%%
&= e(A_0\hat{\textbf{x}}+A_0i\sigma\hat{\textbf{y}})
\hbar v
\sqrt
{
\frac
{2\pi\epsilon^i_{s,0,\eta}}
{\Delta}
}
\begin{pmatrix}
\cos(\frac{\theta_{s,\eta}}{2})
& 
\sin(\frac{\theta_{s,\eta}}{2})e^{i\eta\varphi_{\textbf{p}}}
\end{pmatrix}
\\\nonumber
&\times
(v
\begin{pmatrix}
0&\eta\\
\eta&0
\end{pmatrix}
\hat{\textbf{x}}
+
v
\begin{pmatrix}
0&-i\\
i&0
\end{pmatrix}
\hat{\textbf{y}}
)
\begin{pmatrix}
\frac{\Delta - s\lambda\eta}{(vp)^2+\epsilon^i_{s,0,\eta}(\Delta-s\lambda\eta)}
\\\nonumber
\frac{vp}{(vp)^2+\epsilon^i_{s,0,\eta}(\Delta-s\lambda\eta)}
\end{pmatrix}
\\\nonumber
%%%%%%%%%%%%%%%%%%%%%%% 4 %%%%%%%%%%%%
&=
(evA_0)
(\hbar v)
\sqrt
{
\frac
{2\pi\epsilon^i_{s,0,\eta}}
{\Delta}
}
\Bigg\{
(\eta+\sigma)\frac{v(\textbf{p}-\textbf{k})}{v^2(\textbf{p}-\textbf{k})^2+\epsilon^i_{s,0,\eta}(\Delta-s\lambda\eta)}\cos(\frac{\theta_{s,\eta}}{2})
\\
\label{M_0_appen}
&+
(\eta-\sigma)\frac{\Delta-s\lambda\eta}{v^2(\textbf{p}-\textbf{k})^2+\epsilon_i(\Delta-s\lambda\eta)}\sin(\frac{\theta_{s,\eta}}{2})e^{i\eta\varphi_{\textbf{p}}}
\Bigg\}.
\end{align}
Based on Eq.~\ref{EqCurrentMain}, the current density $j_{0x}$ can be written as
\begin{equation}
\label{j0x_appen}
j_{0x}=\frac{2\pi en_i\tau}{\hbar}\int
\frac{v_x(\textbf{p})d\textbf{p}}{(2\pi\hbar)^2}
|M_0(\textbf{p},\textbf{k})|^2
\delta(E_{c}(\textbf{p})-E_i-\hbar \omega).    
\end{equation}
By plugging Eq.~\ref{M_0_appen} into Eq.~\ref{j0x_appen}, 

\begin{align}
j_{0x} &= \nonumber
\frac{2\pi e n_i \tau}{\hbar}\frac{1}{(2\pi\hbar)^2}\int
\frac{vp}{E_c(\textbf{p})}\cos\varphi_{\textbf{p}} d\textbf{p}
(evA_0)^2
(\hbar v)^2
\frac
{2\pi\epsilon_i}
{\Delta}
\\&\times
\Bigg\{
\frac{(\eta+\sigma)^2v^2(\textbf{p}-\textbf{k})^2\cos^2(\frac{\theta_{s,\eta}}{2})}
{\Big[v^2(\textbf{p}-\textbf{k})^2+\epsilon_i(\Delta-s\lambda\eta)\Big]^2}
+
\frac{(\eta-\sigma)^2(\Delta-\lambda\eta)^2\sin^2(\frac{\theta_{s,\eta}}{2})}
{\Big[\Big(v(\textbf{p}-\textbf{k})\Big)^2+\epsilon_i(\Delta-s\lambda\eta)\Big]^2}
\Bigg\}\delta(E_c(\textbf{p})-E_i-\hbar w).
\end{align}

\noindent
The current density can be split into two contributions $j_{0x}=j^1_{0x} + j^2_{0x}$
\begin{align}
\label{j1_label}
j^1_{0x} = &
\beta'_0\int\frac{vp}{E_\uparrow}\cos\varphi d\textbf{p}
\frac{(\eta+\sigma)^2\Big(v(\textbf{p}-\textbf{k})\Big)^2}{\Big[\Big(v(\textbf{p}-\textbf{k})\Big)^2+\epsilon_i(\Delta-\lambda\eta)\Big]^2}
\cos^2(\frac{\theta_{s,\eta}}{2})
\delta(E_c(\textbf{p}) - E_i - \hbar w),
\\
j^2_{0x} = &
\label{j2_label}
\beta'_0\int\frac{vp}{E_\uparrow}\cos\varphi d\textbf{p}
\frac{(\eta-\sigma)^2(\Delta-\lambda\eta)^2}{\Big[\Big(v(\textbf{p}-\textbf{k})\Big)^2+\epsilon_i(\Delta-\lambda\eta)\Big]^2}
\sin^2(\frac{\theta_{s,\eta}}{2})
\delta(E_c(\textbf{p}) - E_i - \hbar w)
\end{align}

\noindent
The definition of $\cos{\theta_{s,\eta}}$ and $\sin{\theta_{s,\eta}}$ reads
\begin{equation}
\label{cosine}
\cos^2{\frac{\theta_{s,\eta}}{2}}
=
\frac{1+\cos{\theta_{s,\eta}}}{2}
=
\frac{2E_{s,\eta}(\textbf{p}) + (\Delta-s\lambda\eta)}{4E_{s,\eta}(\textbf{p})},
\end{equation}
\begin{equation}
\sin^2{\frac{\theta_{s,\eta}}{2}}
=
\frac{1-\cos{\theta_{s,\eta}}}{2}
=
\frac{2E_{s,\eta}(\textbf{p}) - (\Delta-s\lambda\eta)}{4E_{s,\eta}(\textbf{p})}.
\end{equation}

\noindent
By assumption $p\gg\textbf{k}$, the terms in $j_1$ and $j_2$ can be simplified as
\begin{align}
\frac{1}{\Big[\Big(v(\textbf{p}-\textbf{k})\Big)^2+\epsilon_i(\Delta-\lambda\eta)\Big]^2} &
\approx  
\frac{1}{\Big[(vp)^2+\epsilon_i(\Delta-\lambda\eta)\Big]^2}
\Bigg\{
1
+
\frac{4pv^2\cos\varphi k}{(vp)^2+\epsilon_i(\Delta-\lambda\eta)}
\Bigg\},  &  
\\
\frac{(\textbf{p}-\textbf{k})^2 }{\Big[\Big(v(\textbf{p}-\textbf{k})\Big)^2+\epsilon_i(\Delta-\lambda\eta)\Big]^2}
&\approx\nonumber 
\frac{1}{\Big[(vp)^2+\epsilon_i(\Delta-\lambda\eta)\Big]^2}
\Bigg\{p^2 - 2pk\cos\varphi 
\\&
+
\label{approxi2}
\frac{4p^3v^2\cos\varphi k}{(vp)^2+\epsilon_i(\Delta-\lambda\eta)}
\Bigg\}.  
\end{align}
By substituting Eq.~(\ref{cosine}) - Eq.~(\ref{approxi2}) into Eq.~(\ref{j1_label}) and Eq.~(\ref{j2_label}), we obtain
\begin{align}
j^1_{0x}  = &
\beta'_0\Theta[\delta\hbar\omega_{s,0,\eta}]\frac{k\pi}{v}
\frac{(\eta+\sigma)^2\Big[(\Delta-s\lambda\eta)+\delta\hbar\omega_{s,0,\eta}
\Big]^2}{
\Big[(\delta\hbar\omega_{s,0,\eta})^2 +(\Delta-s\lambda\eta)\hbar \omega\Big]^2
}
\frac
{
\delta\hbar\omega_{s,0,\eta}
}
{(\Delta-s\lambda\eta) + \delta\hbar\omega_{s,0,\eta} }
\\&
\times
\left(\frac{4\Big((\Delta-s\lambda\eta)+\delta\hbar\omega_{s,0,\eta}\Big)\delta\hbar\omega_{s,0,\eta}}{(\delta\hbar\omega_{s,0,\eta})^2 +(\Delta-s\lambda\eta)\hbar \omega}
-2\right),
\\
j^2_{0x}   =&
\beta'_0\Theta[\hbar \omega - \epsilon_i]\frac{k\pi}{v}
\frac{4(\eta-\sigma)^2(\Delta-s\lambda\eta)^2}
{
(\Delta-s\lambda\eta) + \delta\hbar\omega_{s,0,\eta}
}
\frac{
\Big[
(\Delta-s\lambda\eta) + \delta\hbar\omega_{s,0,\eta}
\Big]
(\delta\hbar\omega_{s,0,\eta})^2
}{\Big[(\delta\hbar\omega_{s,0,\eta})^2 +(\Delta-s\lambda\eta)\hbar \omega\Big]^3}.
\end{align},
where 
$\beta_{0}'=en_i\tau v^2\epsilon_i(evA_0)^2/\hbar\Delta$ and $\delta\hbar\omega_{s,m,\eta}=\hbar\omega-\epsilon^i_{s,m,\eta}$. The summation of $j_1$ and $j_2$ gives Eq.~(\ref{EqM0})

% By assuming $p\gg\textbf{k}$,
% \text{let $A = \epsilon_i\Delta+v^2p^2$ and $B = 2pv^2\cos\varphi_\textbf{p}$}

% \begin{align*}
% \frac{1}{(\epsilon_i\Delta+v^2(\textbf{p}-\textbf{k})^2)^2 }
% &\approx
% \frac{1}{(\epsilon_i \Delta+v^2p^2+\xcancel{v^2k^2}-(2pv^2\cos\varphi_\textbf{p}k))^2}
% \\
% &=
% \frac{1}{(A-Bk)^2}
% \\
% &=
% \frac{1}{A^2-2ABk+\xcancel{B^2k^2}}
% \\
% &\approx
% \frac{1}{A^2(1-\frac{2Bk}{A})}
% \\
% &\approx
% \frac{1}{A^2}(1+\frac{2Bk}{A})  
% \end{align*}

% \begin{equation}
% \begin{split}
% (\textbf{p}-\textbf{k})^2
% &\approx p^2+\xcancel{k^2}-2pk\cos\varphi_\textbf{p}
% \end{split}    
% \end{equation}

% Multiplying (9) and (10),
% \begin{equation}
% \begin{split}
% \frac{1}{(\epsilon_i\Delta+v^2(\textbf{p}-\textbf{k})^2)^2 }(\textbf{p}-\textbf{k})^2
% &\approx \frac{1}{A^2}(1+\frac{2Bk}{A})(p^2-2pk\cos\varphi_\textbf{p})
% \\
% &\approx
% \frac{1}{A^2}(p^2-2pk\cos\varphi_\textbf{p}+\frac{2Bp^2}{A}k-\xcancel{\frac{4Bp}{A}cos\varphi_\textbf{p}k^2})
% \\
% &=
% \frac{1}{A^2}(p^2-2pk\cos\varphi_\textbf{p}+\frac{2Bp^2}{A}k)
% \\
% &=
% \frac{1}{(\epsilon_i\Delta+v^2p^2)^2}(p^2+(\frac{4p^3v^2\cos\varphi_\textbf{p}}{\epsilon_i\Delta+v^2p^2}-2p\cos\varphi_\textbf{p})k)
% \end{split}    
% \end{equation}

\subsection{Derivation of $j_{1x}$}  \label{j_1x_appendix}
With the same method to obtain $M_0(\textbf{p},\textbf{k})$, $M_1(\textbf{p},\textbf{k})$ reads
\begin{align}
\label{M_1_appen}
M_1(\textbf{p},\textbf{k}) &= \bra{\psi_{s,\eta}(\textbf{p})}\hat{V}\ket{\chi_{s,1}(\textbf{p}-\textbf{k})}.     
\nonumber
\\\nonumber
&=
(evA_0)
(\hbar v)
\sqrt
{
\frac
{2\pi(\Delta^2 - \lambda^2)}
{\Delta\epsilon^i_{s,1,-1}}
}
\Bigg\{
\frac
{(\sigma-1)\cos(\frac{\theta_{s,-1}}{2})}
{v^2(\textbf{p}-\textbf{k})^2+\epsilon^i_{s,1,-1}(\Delta+s\lambda)}
\\
&-
\frac
{(\sigma+1)v(\textbf{p}-\textbf{k})
\sin(\frac{\theta_{s,-1}}{2})
}
{v^2(\textbf{p}-\textbf{k})^2+\epsilon^i_{s,1,-1}(\Delta+s\lambda)}
e^{-i\varphi_{\textbf{p}}}
\Bigg\}.    
\end{align}
\noindent
Based on Eq.~(\ref{EqCurrentMain}), the current density $j_{1x}$ can be written as
\begin{equation}
\label{j1x_appen}
j_{1x}=\frac{2\pi en_i\tau}{\hbar}\int
\frac{v_x(\textbf{p})d\textbf{p}}{(2\pi\hbar)^2}
|M_1(\textbf{p},\textbf{k})|^2
\delta(E_{c}(\textbf{p})-E_i-\hbar \omega).  
\end{equation}

\noindent
By substituting Eq.~(\ref{M_1_appen}) into Eq.~(\ref{j1x_appen}).
\begin{equation}
\begin{split}
j_{1x} = &
\frac{2\pi e n_i \tau}{\hbar}\frac{1}{(2\pi\hbar)^2}\int
\frac{vp}{E_c(\textbf{p})}\cos\varphi_{\textbf{p}} d\textbf{p}
(evA_0)^2
(\hbar v)^2
\frac
{2\pi(\Delta^2-(\lambda)^2)}
{\Delta\epsilon_i}
\\&\times
\Bigg\{
\frac{(\sigma-1)^2(\epsilon^i_{s,1,-1})^2}{\Big[\Big(v(\textbf{p}-\textbf{k})\Big)^2+\epsilon^i_{s,1,-1}(\Delta+s\lambda)\Big]^2}
\cos^2(\frac{\theta_{s,-1}}{2})
\\&+
\frac{(\sigma+1)^2v^2(\textbf{p}-\textbf{k})^2}{\Big[\Big(v(\textbf{p}-\textbf{k})\Big)^2+\epsilon^i_{s,1,-1}(\Delta+s\lambda)\Big]^2}
\sin^2(\frac{\theta_{s,-1}}{2})
\Bigg\}
\delta(E_c(\textbf{p})-E_i-\hbar w).
\end{split}    
\end{equation}

\noindent
Let $\beta_{1}'=\beta_{0}'[\Delta^2-\lambda^2]/\epsilon_i^2$ and split $j_{1x}$ into $j^1_{1x}$ and $j^2_{1x}$. 
\begin{align}
\label{j11x}
j^1_{1x} = &
\beta'_1\int\frac{vp}{E_c(\textbf{p})}\cos\varphi_{\textbf{p}} d\textbf{p}
\frac{(\eta+\sigma)^2\epsilon_i^2
\cos^2(\frac{\theta_{s,-1}}{2})
}{\Big[\Big(v(\textbf{p}-\textbf{k})\Big)^2+\epsilon_i(\Delta-\lambda\eta)\Big]^2}
\delta(E_c(\textbf{p}) - E_i-\hbar w),
\\
\label{j21x}
j^2_{1x} = &
\beta'_1\int\frac{vp}{E_c(\textbf{p})}\cos\varphi_{\textbf{p}} d\textbf{p}
\frac{(\eta-\sigma)^2v^2(\textbf{p}-\textbf{k})^2\sin^2(\frac{\theta_{s,-1}}{2})}{\Big[\Big(v(\textbf{p}-\textbf{k})\Big)^2+\epsilon_i(\Delta-\lambda\eta)\Big]^2}
\delta(E_c(\textbf{p}) - E_i - \hbar w).
\end{align}

\noindent
By substituting Eq.~(\ref{cosine}) - Eq.~(\ref{approxi2}) into Eq.~(\ref{j11x}) and Eq.~(\ref{j21x}),
\begin{align}
j^1_{1x}  = &\nonumber
\beta'_1\Theta[\delta\hbar\omega_{s,0,\eta}]\frac{k\pi}{v}
\frac{(\sigma+1)^2\Big[(\Delta+s\lambda)+\delta\hbar\omega_{s,1,-1}
\Big]}{
\Big[(\delta\hbar\omega_{s,1,-1})^2 +(\Delta+s\lambda)\hbar \omega\Big]^2
}
\frac
{
(\delta\hbar\omega_{s,1,-1})^2
}
{(\Delta+s\lambda) + 2\delta\hbar\omega_{s,1,-1} }
\\&
\times
\left(\frac{4\Big((\Delta+s\lambda)+\delta\hbar\omega_{s,1,-1}\Big)\delta\hbar\omega_{s,1,-1}}{(\delta\hbar\omega_{s,1,-1})^2 +(\Delta+s\lambda)\hbar \omega}
-2\right).
\\
j^2_{1x}  = &
\beta'_1\Theta[\hbar \omega - \epsilon_i]\frac{k\pi}{v}
\frac{4(\sigma-1)^2(\epsilon^i_{s,1,-1})^2}
{
(\Delta+s\lambda) + 2\delta\hbar\omega_{s,1,-1}
}
\frac{
\Big[
(\Delta+s\lambda) + \delta\hbar\omega_{s,1,-1}
\Big]
(\delta\hbar\omega_{s,1,-1})
}{\Big[(\delta\hbar\omega_{s,1,-1})^2 +(\Delta+s\lambda)\hbar \omega\Big]^3}.
\end{align}

\bibliographystyle{unsrt}
\bibliography{thesis}

% \newpage
\vspace*{3cm}

\end{document}